\documentclass[prd,aps,showpacs,preprintnumbers,nofootinbib,eqsecnum, preprintnumbers]{revtex4}

\usepackage{graphicx}
\usepackage{epsf}
\usepackage{amsmath}
\usepackage{epstopdf}
\usepackage{bm}
\usepackage{color}
\usepackage{tabularx}
\usepackage{enumitem}
\usepackage{float}
\usepackage{array,booktabs}
\usepackage{footnote}
\usepackage{threeparttable}
\usepackage{graphicx}
\usepackage{hyperref}
\usepackage{amssymb,epsf}
\usepackage{latexsym}
\usepackage{epstopdf}
\usepackage{epsfig}
\usepackage{eurosym}
\usepackage{amsfonts}
\usepackage{xcolor}
\usepackage{subfigure}

\begin{document}

\title{Complexity Conjecture of Regular Electric Black Holes}
\author{ B. Bahrami Asl$^{1,2}$\thanks{%
email address: banafsheh.bahrami@shirazu.ac.ir}, S. H. Hendi$^{1,2,3}$%
\thanks{%
email address: hendi@shirazu.ac.ir} and S. N. Sajadi$^{1,2}$\thanks{%
email address: naseh.sajadi@gmail.com} }
\affiliation{$^1$Department of Physics, School of Science, Shiraz
University, Shiraz
71454, Iran \\
$^2$Biruni Observatory, School of Science, Shiraz University,
Shiraz 71454,
Iran \\
$^3$Canadian Quantum Research Center 204-3002 32 Ave Vernon, BC
V1T 2L7 Canada }

\begin{abstract}
Recently, the action growth rate of a variety of four-dimensional
regular magnetic black holes in $\mathcal{F}$ frame is obtained in
\cite{ElMoumni:2020nse}. Here, we study the action growth rate of
a four-dimensional regular electric black hole in $\mathcal{P}$
frame that is the Legendre transformation of $\mathcal{F}$ frame.
We also investigate the action growth rates of the Wheeler-De Witt
patch for such black hole configurations at the late time and
examine the Lloyd bound on the rate of quantum computation. We
show that although the form of the Lloyd bound formula remains
unaltered, the energy modifies due to a non-vanishing trace of the
energy-momentum tensor and some extra terms may appear in the
total growth action. We also investigate the asymptotic behavior
of complexity in two conjectures for static and rotating regular
black holes.
\end{abstract}

\maketitle


\section{Introduction}

AdS/CFT correspondence \cite{AdSCFT1,AdSCFT2,AdSCFT3} is a leading
formalism providing a consistent relation between gauge and
gravitation theories and has given us a rather deep insight on
possible unification of them. Such a correspondence captures an
equivalence between the conformal field theory and its dual theory
of gravity in AdS space with a strong/weak duality point of view.
It is believed that it provides a non-perturbative formulation
access to strongly coupled regimes of quantum field theories which
cannot be accessed by the traditional perturbation approach.
Indeed, AdS/CFT correspondence is one of the important examples of
entanglement between gravity and quantum information
\cite{AdSCFT4,AdSCFT5,AdSCFT6,AdSCFT7}. Complexity has recently
been introduced as a complement to entanglement providing
additional information on the quantum properties of black holes.

In order to calculate the holographic duality of quantum computational
complexity one may follow two proposals. The $complexity = volume$ (CV) \cite%
{A1,A2} and complexity = action (CA) \cite{B1,B2} conjectures. The
former one states that the complexity of the boundary state is
proportional to the maximum volume of codimension hypersurface
bounded by the CFT slices and the latter conjecture specifies that
complexity on the boundary CFT is proportional to the on-shell
action of the Wheeler-DeWitt(WDW) patch in the bulk. According to
these conjectures, Lloyd showed that the growth rate of complexity
for the Schwarzschild like black holes is bounded by the total
energy of the system \cite{Lloyd2000}. Cai, et.al proved that the
so-called
Lloyd bound should be modified for charges or rotating black holes \cite%
{Cai2016}. Although there are many cases of black holes that
confirm the mentioned upper bound \cite{L1,L2,L3,L4,L5,L6,L7}, its
validity is an open question since there are different types of
black hole violating the Lloyd bound \cite{V1,V2,V3}. Accordingly,
the validity of CV and CA conjectures and their possible
modifications deserve further study.

In this direction, here, we focus on the holographic complexity of
nonlinearly charged regular black holes in $\mathcal{P}$ frame.
Indeed, the motivation for studying this black hole is three
folds: regularity, nonlinearity and complexity. The first item is
the regularity of the black hole with known action, exactly. The
second one is nonlinearity which has the key role to avoid the
singularity. Although such a nonlinear electrodynamics is an ad
hoc model near the black hole, it behaves like the Maxwell field
far from the horizon radius. The third one is the complexity of
black holes which is reasonable for regular black holes since the
singularity may violate the interpretation of physical quantities
such as entangled entropy and complexity.

The subject of singularity itself and its possible solution
(leading to regularity) are highly interesting from different
viewpoints: mathematical and geometrical points of view, classical
gravity and its quantum standpoints, cosmological side, effective
field theory and AdS/CFT correspondence features, string theory
and supergravity aspects, etc. Indeed there are two important
singularities that their natures are not yet fully understood: the
singularity that is covered by an event horizon of black holes and
the big bang singularity. So one of the substantial questions is
how one can prevent the singularity. Although it is believed that
quantum gravity can be able to smooth out singularities, there is
no consistent theory of quantum gravity overcoming this issue
despite many attempts. Thus, the question should again be asked
how we can preclude the singularity in an alternative theory.

Regarding the Einstein theory of gravity coupling minimally with
an appropriate nonlinear electrodynamics theory, various regular
black hole metrics are constructed (for an incomplete list, please
take a look at \cite{Balart2014,Ayon-Beato1998,Bronnikov2001,Arellano2006,Hassain2008,Hollenstein2008,Balart2009}%
). The exact regular solution of $f(R)$ gravity has been studied in \cite%
{Hollenstein2008}. Gravitational lensing \cite{Eiroa:2010wm}, dynamical \cite%
{Fernando:2012yw} and thermodynamical stability
\cite{Hendi:2020knv} of regular black hole has been studied. A new
Smarr-type formula for the black hole in nonlinear electrodynamics
has been obtained in \cite{kastor09,Balart:2017dzt}. In this
paper, we concentrate on the complexity of a simple static and
rotating regular black hole with nonlinear electrodynamics as a
source. The rest of the paper is organized as follows. In Sec.
\ref{sec2}, we briefly review the complexity and thermodynamics of
regular black hole, especially deriving its Smarr formula. Section
\ref{sec3} is devoted to study the complexity of
large-extremal-static regular black hole in the framework of CA
and CV conjectures and comparing the results in two frameworks. In
Sec. \ref{sec4}, we repeated the same computation of complexity
for the large-extremal rotating regular black holes. After that,
conclusions are given. In the appendix, we also investigate two
other regular cases and calculated thermodynamic quantities,
complexity and complexity of formation numerically.

\section{basic formalism \label{sec2}}

Electrically charged black hole solutions can be studied through
an alternative form of nonlinear electrodynamics obtained by the
Legendre transformation. The action of nonlinear electrodynamics
in $\mathcal{P}$ frame minimally coupled to the Einstein gravity
with the York-Gibbons-Hawking surface term and the two-dimensional
joint term is given by
\begin{eqnarray}
\mathcal{A}&=&\frac{1}{4\pi}\int d^{4} x \sqrt{-g}\left[ \frac{1}{4} (%
\mathcal{R}-2\Lambda) -\left(\frac{1}{2} P^{\mu\nu} F_{\mu\nu}- \mathcal{H} (%
\mathcal{P})\right)\right]  \notag \\
&&+\frac{1}{8\pi}\int d^{3}x \sqrt{-h} \mathcal{K}+\frac{1}{8\pi}\int d^{2}x
\sqrt{-\gamma}\; a ,  \label{eqI}
\end{eqnarray}
where $g$ is the determinant of the metric, $\mathcal{R} $ is the Ricci
scalar and the negative cosmological constant is denoted by $\Lambda=-\frac{3%
}{l^2}$. In addition, the anti-symmetric tensor $P_{\mu\nu}= \mathcal{L}%
_{F}F_{\mu \nu}$ in which $\mathcal{L}_{\mathcal{F}}\equiv\frac{\partial
\mathcal{L}}{\partial \mathcal{F}} $, $\mathcal{F}=\frac{1}{4}%
F_{\mu\nu}F^{\mu\nu } $ and $F_{\mu \nu} $ is the Faraday tensor.
Furthermore, the structure function is $\mathcal{H(\mathcal{P})}=2\mathcal{F}
\mathcal{L}_{\mathcal{F}}-\mathcal{L}$ with $\mathcal{P} \equiv\frac{1}{4}%
P_{\mu\nu}P^{\mu\nu }=\mathcal{F} \mathcal{L}_{\mathcal{F}}^2$ since $d%
\mathcal{H}=\mathcal{L}_{\mathcal{F}}^{-1}d\left(\mathcal{F}\mathcal{L}_{%
\mathcal{F}}^2\right)=\mathcal{H}_{\mathcal{P}}d\mathcal{P}$, where $%
\mathcal{H}_{\mathcal{P}}=\frac{d\mathcal{H}}{d\mathcal{P}}=\mathcal{L}_{%
\mathcal{F}}^{-1}$. Moreover, $h$ and $\gamma$ are, respectively, the
determinant of the induced metrics $h_{\mu \nu}$ in three-dimensions and
two-dimensional $\gamma_{\mu \nu}$. Also, $\mathcal{K}$ is the trace of the
extrinsic curvature of the induced metric $h_{\mu \nu}$
\begin{equation}  \label{extcur}
\mathcal{K}=n^{\mu}{}_{;\mu}=\dfrac{1}{\sqrt{-g}}\partial_{\mu}(\sqrt{-g}%
n^{\mu}),
\end{equation}
where $n^{\mu}$ is the normal vector. The integrant $a$ is defined as
\begin{equation*}
a=\ln \left(-\frac{1}{2}N.\bar{N}\right),
\end{equation*}
in which $N$ is the future-directed null normal to the left-moving null
surface and $\bar{N}$ denotes the future-directed null normal to the
right-moving null surface.

\begin{center}
\begin{figure}[tbp]
\hspace{0.2cm} \centering
\subfigure[]{\includegraphics[width=0.4\columnwidth]{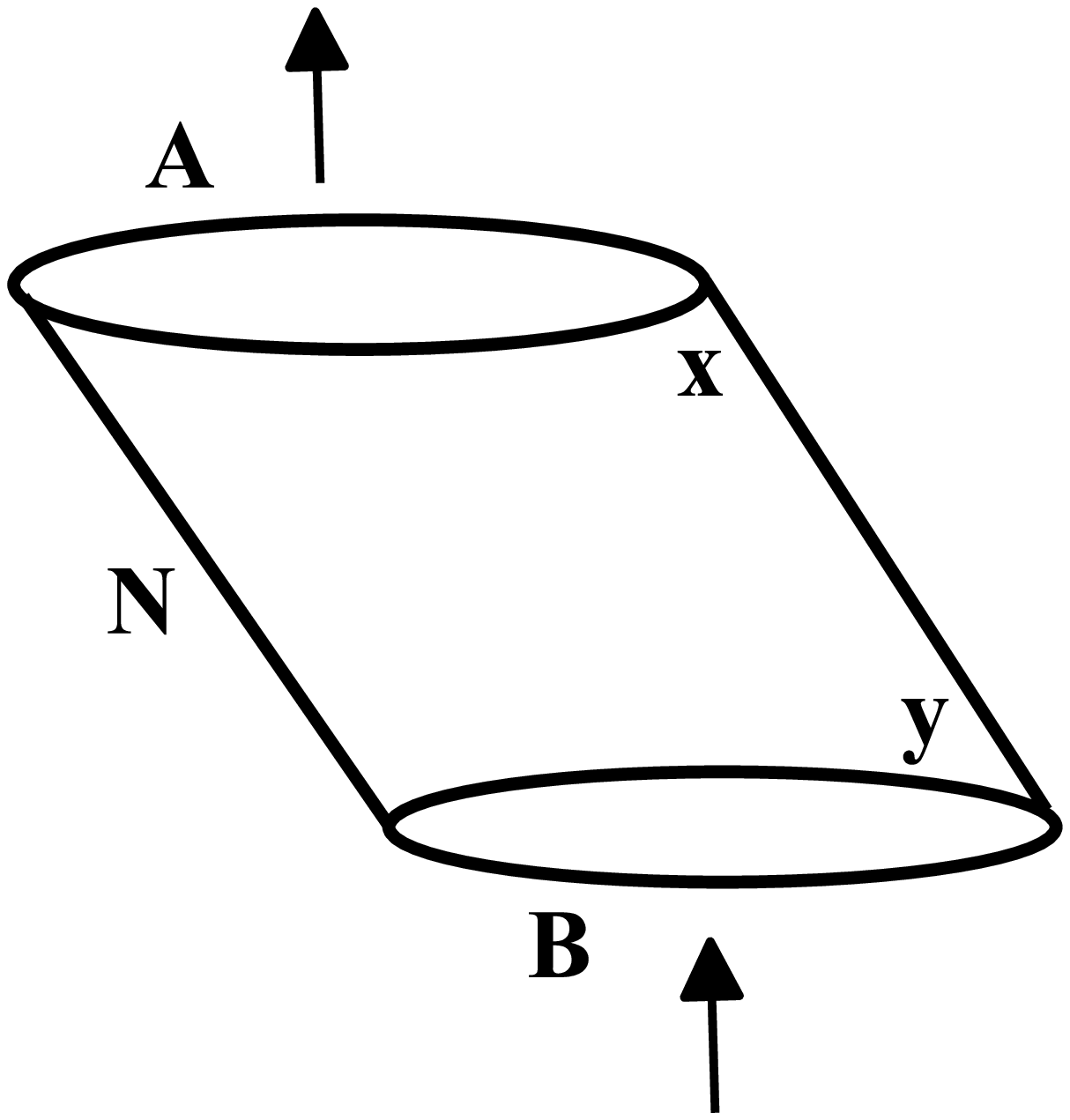}} %
\subfigure[]{\includegraphics[width=0.4\columnwidth]{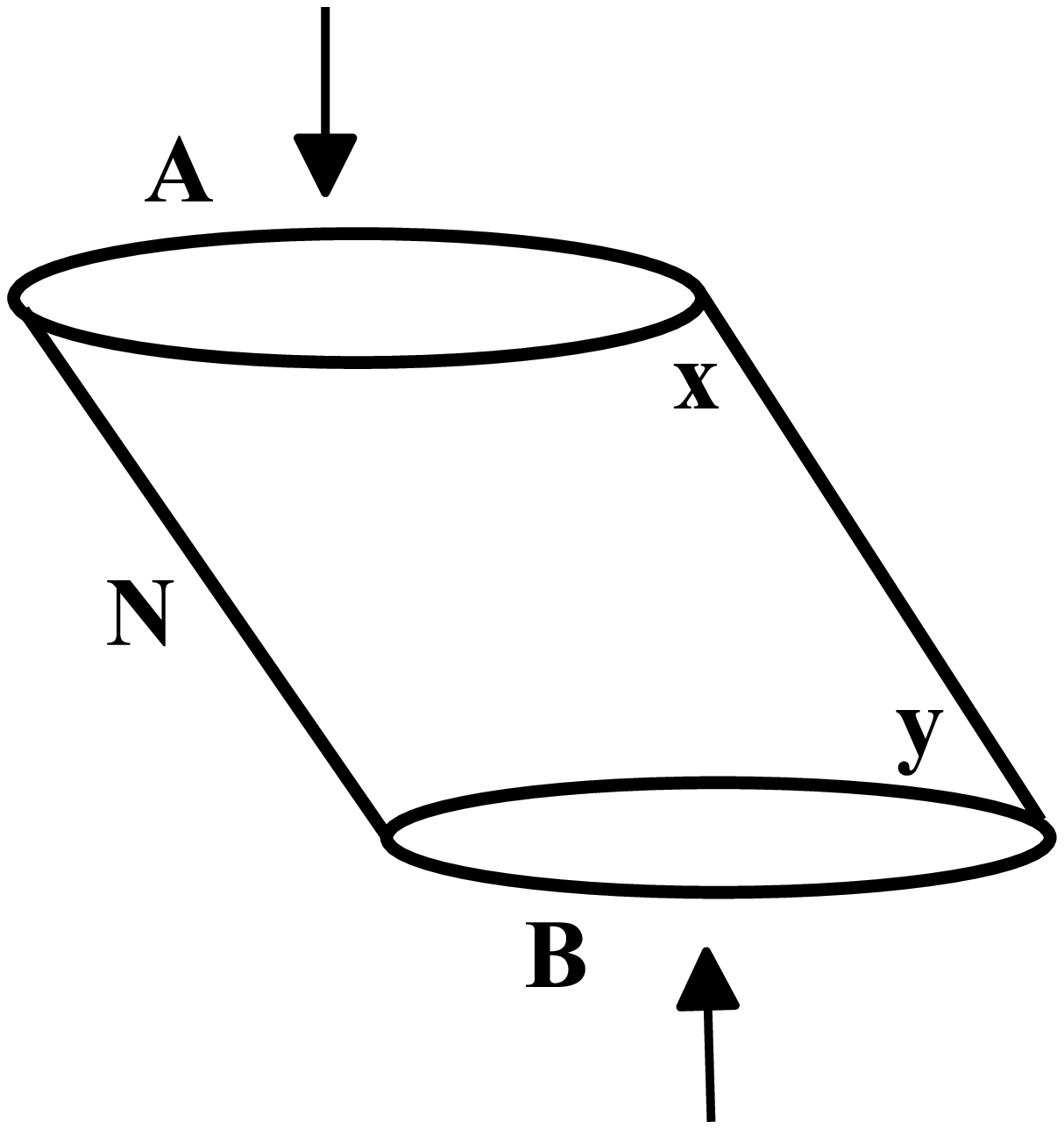}}
\caption{Closed hypersurfaces consisting of a past spacelike
surface $B$, a truncated null cone $N$ and a future spacelike
surface $A$. (a) Same direction of normal vectors, (b) Opposite
direction of normal vectors. } \label{fig1}
\end{figure}
\end{center}

The first integral of Eq. (\ref{eqI}) represents the bulk action while the
second and third ones stand for the boundary and joint parts of the WDW
patch \cite{Lehner:2016vdi}. The WDW patch is used to obtain the rate of
temporal change of the action. A typical WDW patch of a black hole with two
horizons is shown in Fig. \ref{wdw} which is evolved in time from $t_{0}$ to
$t_{0}+\delta t$. According to this figure, the contribution of different
parts is bulk region $V_{1}$ and $V_{2}$, and null-null surface joints $A$, $%
B$, $C$ and $D$ (for more details, see \cite{Lehner:2016vdi}).
Hence, we can write
\begin{eqnarray}
\partial \mathcal{A} &=&\dfrac{1}{4 \pi G}\int_{V_{1}} \sqrt{-g}~\mathcal{L}%
\; dt dr d\theta d\phi -\dfrac{1}{4 \pi G}\int_{V_{2}} \sqrt{-g}~\mathcal{L}%
\; dt dr d\theta d\phi  \notag \\
&&+\frac{1}{8\pi G} \int_{B}\sqrt{-\gamma }\ a_{B}\ d\theta d\phi - \frac{1}{%
8\pi G}\int_{A} \sqrt{-\gamma}\ a_{A}\ d\theta d\phi  \notag \\
&&+\frac{1}{8\pi G}\int_{D} \sqrt{-\gamma}\ a_{D}\ d\theta d\phi -\frac{1}{%
8\pi G}\int_{C}\ \sqrt{-\gamma}\ a_{C}\ d\theta d\phi \; ,  \label{action 2}
\end{eqnarray}
where $\mathcal{L}$ is the Lagrangian of Einstein-nonlinear electrodynamics.
In order to compute the contribution of joints, we use the following
transformation of $N $ and $\bar{N}$
\begin{equation}
N_{\alpha }=-b_{1}\partial _{\alpha }(t-r^{\ast }),\ \ \ \ \ \ \bar{N}
=b_{2}\partial _{\alpha }(t+r^{\ast }),
\end{equation}
in which $b_1$ and $b_2$ are two arbitrary positive constants and the
tortoise coordinate $r^{\ast}$ is defined as $r^{\ast }=\int \frac{dr}{f(r)}$%
. So, the contribution of joints is calculated as
\begin{equation}
S_{B}-S_{A}=\frac{\delta t}{4}\ \left(2 r f(r) \left[ \ln \left(\frac{f(r)}{%
b_{1}b_{2}}\right) + \frac{r f^{\prime }(r)}{2 f(r)} \right]\right) ^{r_{B}}
_{r_{A}},  \label{joint1}
\end{equation}
\begin{equation}
S_{D}-S_{C}=\frac{\delta t}{4}\ \left(2 r f(r) \left[ \ln \left(\frac{f(r)}{%
b_{1}b_{2}}\right) + \frac{r f^{\prime }(r)}{2 f(r)}
\right]\right)_{r_{A}},  \label{joint12}
\end{equation}
where prime denotes derivative with respect to $r$. In Fig.
\ref{fig1}, we assume close cylindrical hypersurfaces consisting
of a past spacelike surface $B$, a truncated null cone $N$, a
future spacelike surface $A$ and boundaries for intersection of
spacelike and null surfaces $x$ and $y$.

The contribution of surfaces $A$ and $B$ canceled each other in
Fig. (\ref{fig1}a) due to the same direction of normal vectors.
While for Fig. (\ref{fig1}b) according to opposite direction of
normal vectors, contributions of boundaries of intersection of
surfaces, $x$ and $y$ cancel each other. Therefore depending on
what we choose for direction of normal vector, boundary terms or
joint contributions will be vanished \cite{Lehner:2016vdi}.

It is notable that applying the variational principle to the
action (\ref{eqI}), one finds the field equations as
\begin{equation}  \label{gg}
G_{\mu}^\nu+\Lambda\delta_{\mu}^\nu=T_{\mu}^\nu= \frac{1}{2}\left[\mathcal{H}%
_{\mathcal{P}}P_{\mu\lambda}P^{\nu\lambda}- \delta_{\mu}^\nu\left(2\mathcal{P%
}\mathcal{H}_{\mathcal{P}} -\mathcal{H}\right)\right],
\end{equation}
\begin{equation}  \label{nn}
\nabla_\nu P^{\mu\nu}=0,
\end{equation}
where $G_{\mu \nu} $ is the Einstein tensor. By integrating equation (\ref%
{nn}) with the assumption of spherical symmetry spacetime, we obtain the
following static solution
\begin{equation}  \label{eqmetric1}
ds^2=-f(r)dt^2+\frac{dr^2}{f(r)}+r^2d\Omega_{2}^2,\hspace{0.5cm}f(r)=1-\frac{%
1}{r}\int \rho(r)r^2 dr,
\end{equation}
\begin{equation}  \label{eqn:p}
P_{\mu \nu}=2\delta^{t}_{[\mu}\delta^{r}_{\nu]} \frac{q}{r^{2}},\hspace{0.5cm%
} \mathrm{or} \hspace{0.5cm} \mathcal{P}=-\frac{q^{2}}{2r^{4}},
\end{equation}
where $q$ is an integration constant and $\mathcal{H}=2\rho(r)$.

In the following, for the importance of Smarr relation, we will
obtain it for the case of spherically symmetric static charged
regular black holes. In order to do that we use the Komar formula
for the mass of black holes as follows
\begin{equation}
M=-\dfrac{1}{8\pi }\oint_{\infty }\nabla ^{\alpha }\xi ^{\beta }dS_{\alpha
\beta },  \label{eqkomar}
\end{equation}%
where $\xi $ is a timelike Killing vector which satisfies Killing equation
and $dS_{\alpha \beta }$ is a two-dimensional surface element of
the boundary at infinity. Since the spacetime has two boundaries,
we will write the Komar integral for the mass as a sum of an
integral over a closed null surface at the horizon $H$ and an
integral on the spacelike hypersurface $\Sigma $ which is bounded
by the horizon and infinity as
\begin{equation}
M=-\dfrac{1}{8\pi }\oint_{H}\nabla ^{\alpha }\xi ^{\beta }dS_{\alpha \beta }-%
\dfrac{1}{8\pi }\int_{\Sigma }\nabla_{\beta} \nabla ^{\alpha }\xi
^{\beta }d\Sigma _{\alpha }.  \label{eqkomarr}
\end{equation}%
For the first term, we have
\begin{equation}
-\dfrac{1}{8\pi }\oint_{H}\nabla ^{\alpha }\xi ^{\beta }dS_{\alpha \beta }=%
\dfrac{\kappa A}{4\pi },
\end{equation}%
where $A$ is the horizons surface area, $\kappa $ denotes the
surface gravity that is constant at the event horizon and
satisfies $\xi ^{\alpha }{}_{;\beta }\xi _{\alpha }=\kappa \xi
_{\beta }$. In order to evaluate the second term in equation
(\ref{eqkomarr}), we consider the Stokes theorem for an
antisymmetric tensor field $B^{\alpha \beta }$ as follows
\begin{equation}
\oint_{S}B^{\alpha \beta }dS_{\alpha \beta }=2\int_{\Sigma }B^{\alpha \beta
}{}_{;\beta }d\Sigma _{\alpha }.
\end{equation}%
For the antisymmetric tensor $B^{\alpha \beta }=\nabla ^{\alpha }\xi ^{\beta
}$ of the bulk term $\Sigma $, we have
\begin{equation}
B^{\alpha \beta }{}_{;\beta }=(\nabla ^{\alpha }\xi ^{\beta })_{;\beta
}=-(\nabla ^{\beta }\xi ^{\alpha })_{;\beta }=-\square \xi ^{\alpha },
\end{equation}%
where $\square =\nabla _{\alpha }\nabla ^{\alpha }$. Recalling that $\nabla
_{\rho }\nabla _{\mu }\xi _{\nu }=R_{\nu \mu \rho }{}^{\sigma }\xi _{\sigma
} $, we get
\begin{equation}
\oint_{S}\nabla ^{\alpha }\xi ^{\beta }dS_{\alpha \beta
}=-2\int_{\Sigma }R^{\alpha }{}_{\beta }\xi ^{\beta }d\Sigma
_{\alpha }.
\end{equation}%
Now, by using the Einstein field equation $R_{\alpha \beta }=4\pi \left(
2T_{\alpha \beta }-g_{\alpha \beta }T\right) $, we obtain
\begin{equation}
\oint_{S}\nabla ^{\alpha }\xi ^{\beta }dS_{\alpha \beta }=-16\pi
\int_{\Sigma }\left( T_{\beta }^{\alpha }-\dfrac{1}{2}\delta _{\beta
}^{\alpha }T\right) \xi ^{\beta }d\Sigma _{\alpha }.  \label{eqmass}
\end{equation}%
On the other hand, considering equation (\ref{gg}), the nonzero components
of energy-momentum tensor are
\begin{equation}
T^{t}{}_{t}=T^{r}{}_{r}=-\dfrac{\mathcal{H}}{8\pi },\hspace{0.5cm}T^{\theta
}{}_{\theta }=T^{\phi }{}_{\phi }=\dfrac{1}{8\pi }\left( 2\mathcal{P}%
\mathcal{H_{P}}-\mathcal{H}\right).
\end{equation}%
By rewriting the RHS of Eq. (\ref{eqmass}), one can obtain
\begin{eqnarray}
-16\pi \int_{\Sigma }\left( T_{\beta }^{\alpha }-\dfrac{1}{2}\delta _{\beta
}^{\alpha }T\right) \xi ^{\beta }d\Sigma _{\alpha } &=&  \notag \\
-16\pi \int_{\Sigma }\left( T_{\beta }^{\alpha }-\dfrac{1}{4}\delta _{\beta
}^{\alpha }T\right) \xi ^{\beta }d\Sigma _{\alpha }+4\pi \int_{\Sigma
}T\delta ^{\alpha }{}_{\beta }\xi ^{\beta }d\Sigma _{\alpha } &=&  \notag \\
-2\int_{\Sigma }\dfrac{q}{r^{2}}Ed\Sigma _{\alpha }+4\pi \int T\sqrt{-h}%
drd\theta d\phi .&&
\end{eqnarray}%
since the timelike Killing vector is $\xi ^{\beta }=\delta
_{t}^{\beta }$ and $E^{2}=-2\mathcal{P}\mathcal{H_{P}}^{2}$ with
$E=q/r^{2}\mathcal{H_{P}}$ identically. Finally, by inserting
above results in Eq. (\ref{eqkomarr}), we obtain
\begin{equation}
M=\dfrac{\kappa A}{4\pi }+q\Phi -\dfrac{1}{2}\int
T\sqrt{-h}drd\theta d\phi,
\end{equation}%
where $h$ is the trace of the induced metric of spacelike hypersurface.

\section{The complexity growth of regular electric black holes \label{sec3}}

Here, we consider a class of known regular electric black holes in
AdS spacetime and calculate its complexity growth rate. For the
sake of completeness, we will point out other regular black hole
solutions in the appendix.


\begin{center}
\begin{figure}[tbp]
\hspace{0.4cm} \centering
\subfigure{\includegraphics[width=0.8\columnwidth]{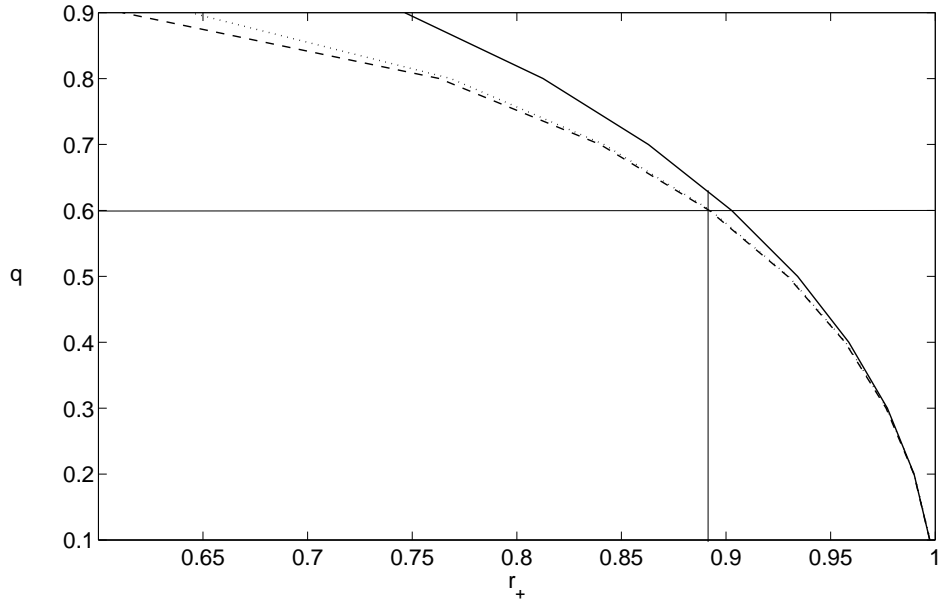}}
\caption{The behavior of q in terms of $r_{+}$ for $M=l=1$. dotted
line (first case), dashed line (second case) and black line (third
case).} \label{r2}
\end{figure}
\end{center}

The motivations of considering the forthcoming kind of
$\mathcal{H}$ function are its simplicity and its agreement to the
correspondence which indicates such a modified nonlinear theory
should reduce to the usual Maxwell theory in the weak-field limit.
The ansatz of structure function which is inspired by the
exponential distribution is defined as \cite{Balart:2014cga}
\begin{equation}  \label{eqhamilton}
\mathcal{H}=\mathcal{P}e^{-\left( \frac{\gamma^{\frac{3}{2}}(-\mathcal{P})^{%
\frac{1}{4}}}{2^{\frac{5}{4}} \chi}\right) }+\frac{6}{l^2},
\end{equation}
in which for the weak field limit ($\mathcal{P} \ll 1$), it
reduces to the Maxwell field as
\begin{equation}
\mathcal{H}\approx \mathcal{P}+\frac{6}{l^2}+\mathcal{O}(\mathcal{P}^{\frac{5%
}{4}}).
\end{equation}

Considering the mentioned structure function with Eqs. (\ref{eqn:p}) and (%
\ref{eqmetric1}), we can find the following spherically symmetric
exponentially black hole solution \cite{Balart:2014cga,
Balart:2014jia, Culetu:2014lca, Hendi:2020knv, Ghosh:2014pba}
\begin{equation}  \label{eqmetric0}
-g_{tt}=f(r)=1+\frac{r^{2}}{l^{2}}-\frac{2\chi}{r}e^{\left(-\frac{\gamma^{2}%
}{2\chi r}\right) }.
\end{equation}

Using the series expansion of this solution for large values of
$r$, it is noticeable that its asymptotical behavior can be found
by the following expression
\begin{equation}
-g_{tt}\approx 1+\frac{r^{2}}{l^{2}}-\frac{2\chi}{r}+\frac{\gamma^{2}}{r^{2}%
}+\mathcal{O}\left(\frac{1}{r^{4}}\right),
\end{equation}
which is the Reissner-Nordstr\"{o}m-AdS metric function provided
$\chi$ and $\gamma$ are associated with the mass and electric
charge of the system, respectively. In other words, the
asymptotical behavior of the obtained solution is completely
matched to the Reissner-Nordstr\"{o}m-AdS black hole.

In order to check the first law and the Smarr formula in the
extended phase space, it is convenient to consider the
cosmological constant as a varying
thermodynamic quantity and interpret it as a thermodynamic pressure, as $P=%
\frac{3}{8\pi l^{2}}$. In addition, its conjugate variable of the introduced
pressure is the thermodynamic volume
\begin{equation}  \label{eqvol}
V=\frac{4\pi}{3}r_{+}^{3},
\end{equation}
where $r_{+}$ is the event horizon radius obtained via $f(r_{+}) = 0$. The
temperature would be found straightforwardly through the use of surface
gravity ($\kappa$) interpretation with the following explicit relation
\begin{equation}
T=\frac{\kappa}{2 \pi}=\left. \frac{1}{4 \pi}\frac{df(r)}{dr}%
\right|_{r=r_{+}}= \left(m-\frac{q^{2}}{2r_{+}}\right) \frac{e^{-\left(
\frac{q^2}{2mr_{+}}\right)}} {2\pi r_{+}^{2}}+\frac{r_{+}}{2\pi l^{2}},
\end{equation}
in which for the large black hole event horizon ($r_{+} \gg l$) becomes
\begin{equation}
T\approx \dfrac{3r_{+}}{4\pi l^2}+\mathcal{O}\left(\dfrac{1}{r_{+}}\right).
\end{equation}
Since we are working in the Einstein gravity, the black hole
entropy $S$ pursues the area law, yielding
\begin{equation}
S=\frac{A}{4}=\pi r_{+}^{2}.
\end{equation}
In addition, the electrostatic potential $\Phi$ can be obtained at the event
horizon versus spatial infinity as the reference
\begin{equation}
\Phi = \int_{r_{+}}^{\infty}E dr=\frac{3m}{2q}-\frac{(6mr_{+}-q^{2})e^{-%
\left(\frac{q^2}{2mr_{+}}\right)}}{4qr_{+}},
\end{equation}
and the asymptotic limit becomes
\begin{equation}
\Phi \approx \dfrac{q}{r_{+}}+\mathcal{O}\left(\dfrac{1}{r_{+}^{5}}\right).
\end{equation}
For a black hole embedded in AdS spacetime, employing the relation
between the cosmological constant and thermodynamic pressure would
result to interpret the mass of black hole as the enthalpy. The
enthalpy can be written in terms of thermodynamic quantities as
\begin{equation}
H=m=\frac{q^2}{2r_{+}\mathcal{W}\left(\frac{q^2 l^2}{r_{+}^2(l^2+r_{+}^2)}%
\right)},
\end{equation}
where $\mathcal{W}$ is the $Lambert\ W$ function. In the case of large black
hole event horizon, we have
\begin{equation}  \label{eqappH}
H\approx \dfrac{r_{+}^{3}}{2 l^{2}}+\dfrac{r_{+}}{2}+\mathcal{O}\left(\dfrac{%
1}{r_{+}}\right).
\end{equation}

It is straightforward to check that these thermodynamic quantities satisfy
the first law of black hole thermodynamics in the enthalpy representation
\begin{equation}
dH=TdS+VdP+\Phi dq,
\end{equation}
and the Smarr relation is given by
\begin{equation}  \label{smarr1}
\frac{H}{2}+PV-TS-\frac{q\Phi}{2}+\frac{1}{4}\int w dv=0,
\end{equation}
where the last term of Eq. (\ref{smarr1}) comes from the fact that the
energy-momentum tensor is not traceless, as
\begin{equation}  \label{eqtrace}
\int w dv=\frac{1}{2}\int_{r_{+}}^{\infty}T^{\mu}_{\mu}\ r^{2}dr = m-\left(m+%
\frac{q^{2}}{2r_{+}}\right)e^{-\left(\frac{q^2}{2mr_{+}}\right)},
\end{equation}
and in the case of $r_{+} \gg l$ becomes
\begin{equation}
\int w dv \approx \dfrac{q^4 l^2}{4 r_{+}^{5}}+\mathcal{O}\left(\dfrac{1}{%
r_{+}^{7}}\right).
\end{equation}

\begin{center}
\begin{figure}[tbp]
\centering
\par
\includegraphics[width=8cm]{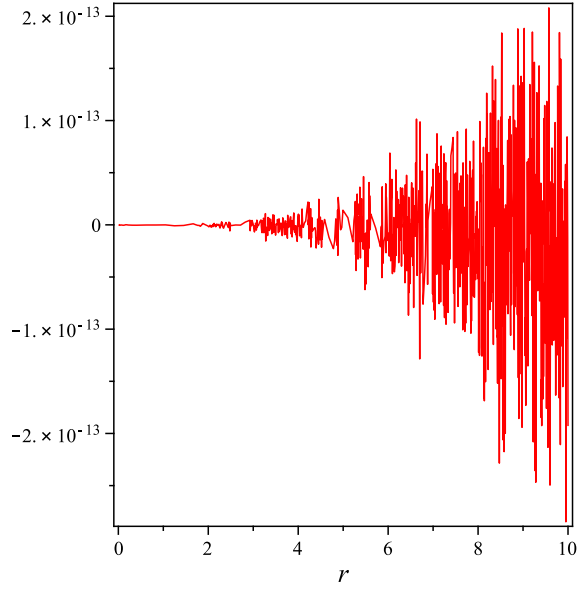}
\caption{The left hand side of Eq. (\protect\ref{smarr1}) versus
$r$ for $M=0.8, q=0.1, l=1$ .} \label{smarr}
\end{figure}
\end{center}

Now, we want to calculate the complexity growth rate in the above
background. For this aim, we should calculate the Ricci scalar for the
solution (\ref{eqmetric0}) as
\begin{equation}  \label{ricciscalar}
\mathcal{R}=-\frac{12}{l^{2}}+\frac{q^{4}e^{-\left( \frac{q^{2}}{2mr}\right)
}}{2mr^{5}},
\end{equation}
whereas by using of Eq. (\ref{eqhamilton}), the Lagrangian of nonlinear
electrodynamics is given by
\begin{equation}  \label{lagrangian}
\mathcal{L}=\frac{-q^2(4mr-q^2)e^{-\left( \frac{q^2}{2mr}\right) }}{2mr^{5}}-%
\frac{6}{l^{2}}.
\end{equation}

Using Eqs. (\ref{ricciscalar}) and (\ref{lagrangian}), we can calculate the
bulk action growth as
\begin{equation}
\frac{d\mathcal{A}_{bulk}}{dt} = \frac{1}{4}\int_{r_{-}}^{r_{+}}r^{2} \left(%
\mathcal{R}+ \frac{6}{l^{2}} -\mathcal{L}\right)dr =m\left[e^{\left( -\frac{%
q^{2}}{2mr}\right) }\right]_{r_{-}}^{r_{+}},
\end{equation}
while the growth rate of the surface term (Eq. (\ref{eqI})) is given by \cite%
{ElMoumni:2020nse}
\begin{eqnarray}
\frac{d\mathcal{A}_{boundary}}{dt}&=&\frac{1}{8\pi}\int_{\partial \mathcal{M}%
}(\sqrt{-h}\mathcal{K})d\Omega_{2}= \frac{1}{2}\left[ \sqrt{f(r)}\frac{%
\partial}{\partial r}(r^{2}\sqrt{f(r)})\right]_{\partial \mathcal{M}} \\
&=&\left[ \frac{3r^{3}}{2l^{2}} +r-\left(\frac{3m}{2} +\frac{q^{2}}{4r}%
\right) e^{\left(-\frac{q^{2}}{2mr}\right) }\right]_{r_{-}}^{r_{+}},  \notag
\end{eqnarray}
where we should note that in the calculation of boundary term we have used $%
\sqrt{-h}=\sqrt{f}r^{2}\sin(\theta)$ and $n^{\mu}=(0,\sqrt{f},0,0)$.

Besides, the contribution of joint terms can be calculated as
\begin{equation}
S_{D}-S_{C}=\frac{\delta t}{4}\ \left(2 r f(r) \left[ \ln \left(\frac{f(r)}{%
b_{1}b_{2}}\right) + \frac{r f^{\prime }(r)}{2 f(r)}
\right]\right) _{r_{C}}.  \label{joint2}
\end{equation}
After some manipulations, we can find
\begin{align}
S_{B}-S_{A}&=  \notag \\
&\frac{\delta t}{4} \left[2me^{-(\frac{q^2}{2mr})} -\frac{q^{2}e^{-(\frac{q^2%
}{2mr})}}{r} +\frac{2r^3}{l^2}+\left(2r-4m e^{-(\frac{q^2}{2mr})} +\frac{2r^3%
}{l^2}\right) \ln\left(\frac{1-\frac{2m}{r} e^{-(\frac{q^2}{2mr})}+ \frac{r^2%
}{l^2}}{b_{1} b_{2}}\right) \right]_{r_A},  \label{eqjoint1} \\
S_{D}-S_{C}&=  \notag \\
&\frac{\delta t}{4}\left[2me^{-(\frac{q^2}{2mr})} -\frac{q^{2} e^{-(\frac{q^2%
}{2mr})}}{r} +\frac{2r^3}{l^2}+ \left(2r-4m e^{-(\frac{q^2}{2mr})} +\frac{%
2r^3}{l^2}\right) \ln\left(\frac{1-\frac{2m}{r} e^{-(\frac{q^2}{2mr})} +%
\frac{r^2}{l^2}}{b_{1}b_{2}} \right)\right]_{r_C}.
\label{eqjoint2}
\end{align}
Since at the late time $r_{A}$ and $r_{C}$ approach to $r_{-}$ and $r_{+}$,
respectively, while $f(r)$ vanishes, one finds Eqs. (\ref{eqjoint1}) and (%
\ref{eqjoint2}) reduce to
\begin{eqnarray}
S_{B}-S_{A}&=&\frac{\delta t}{4}\left[2me^{-\left(\frac{q^2}{2mr}\right)} -%
\frac{q^{2}e^{-\left(\frac{q^2}{2mr}\right)}}{r}+\frac{2r^3}{l^2}\right]%
_{r_A}, \\
S_{D}-S_{C}&=&\frac{\delta t}{4}\left[2me^{-\left(\frac{q^2}{2mr}\right)} -%
\frac{q^{2}e^{-\left(\frac{q^2}{2mr}\right)}}{r}+\frac{2r^3}{l^2}\right]%
_{r_C}.
\end{eqnarray}

So, the total growth rate of the action for such a black hole configuration
within WDW patch at late time approximation is simplified as
\begin{equation}
\frac{d\mathcal{A}}{dt}=(r_{+}-r_{-})\left[ 1+\frac{3}{2l^{2}}%
(r_{+}^{2}+r_{-}^{2}+r_{+}r_{-})\right] +\left( \frac{q^{2}}{4r_{-}}+\frac{m%
}{2}\right) e^{\left( -\frac{q^{2}}{2mr_{-}}\right) }-\left( \frac{q^{2}}{%
4r_{+}}+\frac{m}{2}\right) e^{\left( -\frac{q^{2}}{2mr_{+}}\right) }.
\label{eqtotalaction}
\end{equation}

In order to describe Eq. (\ref{eqtotalaction}) in terms of $r_{-}$ and $%
r_{+} $, we can use the redefinitions $m$ and $q^{2}$ in terms of $r_{-}$
and $r_{+}$ as%
\begin{equation}
m =\frac{r_{-}^{2}(r_{-}^{2}+l^{2})\ln (A)A^{\left( \frac{r_{+}}{%
r_{+}-r_{-}}\right) }}{2(r_{+}-r_{-})l^{2}\mathcal{W}\left( \frac{\ln
(A)A^{\left( \frac{r_{+}}{r_{+}-r_{-}}\right) }r_{-}^{2}(l^{2}+r_{-}^{2})}{%
r_{+}(r_{+}-r_{-})(l^{2}+r_{+}^{2})}\right) },\hspace{0.8cm}
q^{2} =\frac{\ln (A)r_{-}^{2}(l^{2}+r_{-}^{2})r_{+}}{(r_{+}-r_{-})l^{2}e^{%
\frac{r_{+}\ln (A)}{r_{-}-r_{+}}}},
\end{equation}%
where
\begin{equation}
A=\frac{r_{+}(l^{2}+r_{+}^{2})}{r_{-}(l^{2}+r_{-}^{2})}.
\end{equation}%
Although it is straightforward to rewrite Eq. (\ref{eqtotalaction}) in terms
of $r_{-}$ and $r_{+}$, we ignore its explicit relation for the sake of
brevity. Finally, by using the above calculated thermodynamic quantities, it
is easy to show that%
\begin{equation}
\frac{d\mathcal{A}}{dt}\leq 2\left(m-q\phi +\frac{1}{2}\int wdv\right)_{+}-2\left(m-q\phi +%
\frac{1}{2}\int wdv\right)_{-},  \label{rate}
\end{equation}%
indicating that the action growth rate of the black hole in the WDW patch
has been bounded. For instance, for the values $m=0.8,q=0.1,l=1$ this bound
is about $1.5761$, and the differences between left and right (right mines
left) is about $0.0396$ which is clearly non-negative. It is worth
mentioning that although there are some extra terms in the right hand side
of Eq. (\ref{rate}) due to considering the nonlinear electrodynamics as a
source, the action growth rate is bounded.

\subsection{Extreme case: $r_{+}\approx r_{-}$}

Now, we focus on the extremal solutions. By introducing $\alpha=l/r_{+}$, $%
\epsilon=1-r_{-}/r_{+}$ and in the case of extremal, large black hole,
equation (\ref{eqtotalaction}) becomes
\begin{equation}
\dfrac{d \mathcal{A}}{dt}\approx\left (\dfrac{9 r_{+}} {4 \alpha^2}%
-r_{+}\alpha^{2}+\dfrac{7r_{+}}{4}\right)\epsilon + \mathcal{O}(\epsilon^2)+%
\mathcal{O}(\alpha^4),
\end{equation}
in terms of $r_\pm$ leads to
\begin{equation}
\dfrac{d \mathcal{A}}{dt}\approx \dfrac{9 }{4 l^{2}}
(r_{+}^{3}-r_{-}r_{+}^{2})+\dfrac{7}{4}(r_{+}-r_{-})+ \mathcal{O}\left(%
\dfrac{1}{r_{+}}\right).
\end{equation}
By using of equations (\ref{eqvol}), (\ref{eqappH}) and ($r_{+}\approx r_{-}$%
) one can rewrite above equation as follows \cite{Couch:2016exn}
\begin{equation}
\dfrac{d \mathcal{A}}{dt}\approx \dfrac{9}{2}P (V_{+}-V_{-}),\hspace{1cm}%
r_{+}\gg l,
\end{equation}
which shows that in the large black holes the rate of complexity is
proportional to $P\Delta V$, i.e, is controlled by thermodynamical volume.
Besides, it is notable that one can obtain such a relation in terms of
entropy for the static black holes since in the static case the
thermodynamic volume is not independent of entropy ($V=4/(3\sqrt{\pi}%
)S^{3/2} $).

In the case of extremal black hole with small degenerate horizon radius, we
have
\begin{equation}
\dfrac{d \mathcal{A}}{dt}\approx \left(\dfrac{13}{\alpha^{2}} +3\right)%
\dfrac{\epsilon}{4}+\mathcal{O} \left(\dfrac{1}{\alpha^{3}}\right)+\mathcal{O%
}(\epsilon^{2}).
\end{equation}
By using of CV conjecture, the rate of complexity is \cite{An:2018dbz}
\begin{equation}
\dfrac{d\mathcal{C}_{\mathcal{V}}}{dt}=\dfrac{4\pi}{Gl}\sqrt{-f(r_{min})}%
r_{min}^{2},
\end{equation}
where $r_{min}$ is the turning point of maximal surface. The late time limit
of $d\mathcal{C}_{\mathcal{V}}/dt$ becomes
\begin{equation}
\dfrac{d\mathcal{C}_{\mathcal{V}}}{dt}=\dfrac{4\pi}{Gl} \sqrt{-f(\hat{r}%
_{min})}\hat{r}^{2}_{min}, \hspace{1cm}t\longrightarrow\infty,
\end{equation}
where $\hat{r}_{min}$ is the extreme value point of $\sqrt{-f(r_{min})}%
r_{min}^{2}$. For the present case we need to obtain the zeros of the
following equation
\begin{equation}
l^2(6 M \hat{r}_{min}+q^2)\exp\left(-\dfrac{q^2}{2 M \hat{r}_{min}}\right)-2
\hat{r}^{2}_{min}(2 l^2+3 \hat{r}^{ 2}_{min})=0.
\end{equation}
in the case of small $q$, it becomes
\begin{equation}
\hat{r}_{min}\approx r_{0 min}+\varepsilon r_{1 min}=r_{0}+ \dfrac{q^2
l^2\varepsilon }{3M l^{2}-12 r_{0}^{3}-4r_{0}l^{2}},
\end{equation}
where
\begin{equation}
r_{0}=\dfrac{(108Ml^2+4\sqrt{32l^4+729l^4M^2})^{\frac{1}{3}}}{6}- \dfrac{%
4l^2}{(108Ml^2+4\sqrt{32l^4+729l^4M^2})^{\frac{1}{3}}}.
\end{equation}

In the case of extremal black hole with large horizon radius, we can write
\begin{equation}
\dfrac{\hat{r}_{min}}{r_{+}} \approx 1.5-0.5 \varepsilon+ \mathcal{O}%
(\varepsilon,\alpha^{2}),
\end{equation}
and correspondingly, for the late time of growth of complexity, we obtain
\begin{equation}
\dfrac{d\mathcal{C}_{\mathcal{V}}}{dt}\approx \dfrac{40\pi}{\alpha^2}%
(r_{+}-r_{-})\approx 20 P\Delta V.
\end{equation}
The comparison of the complexity growth from the CV and CA dualities are
\begin{equation}\label{Rrate}
\mathcal{R}_{rate}=\dfrac{d\mathcal{C}_{\mathcal{A}}/dt}{d\mathcal{C} _{%
\mathcal{V}}/dt}=\dfrac{9 P \Delta V}{40 P \Delta V}=0.225.
\end{equation}

\subsection{Complexity of formation}

The complexity of formation is the difference between complexity
in the process of forming the entangled TFD state and preparing
two individual copies of the vacuum state of the left and right
boundary CFTs \cite{Chapman:2016hwi,Cottrell:2017ayj}
\begin{equation}
\Delta \mathcal{C}_{\mathcal{A}}=\dfrac{1}{\pi }(\mathcal{A}_{BH}-2\mathcal{A%
}_{AdS}).
\end{equation}

\begin{figure}[tbp]
\centering
\includegraphics[width=6cm] {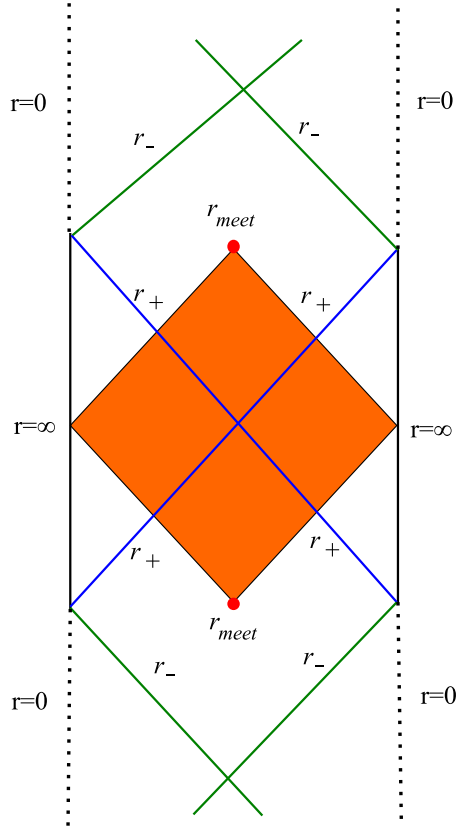}
\caption{Penrose diagram for charged black holes.}
\label{wdw}
\end{figure}
For the case of charged black hole with two horizons, the complexity of
formation in CA conjecture becomes \cite{Chapman:2016hwi}
\begin{equation}\label{eqdeltaca}
\Delta \mathcal{C}_{\mathcal{A}}=\dfrac{1}{\pi }(\Delta \mathcal{A}_{bulk}+%
\mathcal{A}_{joint,meet}).
\end{equation}%
Now, we evaluate the action for the obtained black hole solutions. The
tortoise coordinate for the both horizons of the solution is%
\begin{equation}
r^{\ast }(r)=\dfrac{l^{2}r_{+}\ln \left( \dfrac{|r-r_{+}|}{r+r_{+}}\right) }{%
3r_{+}^{2}+l^{2}-(l^{2}+r_{+}^{2})\mathcal{W}\left( \dfrac{l^{2}q^{2}}{%
r_{+}^{2}(l^{2}+r_{+}^{2})}\right) }+\dfrac{l^{2}r_{-}\ln \left( \dfrac{%
|r-r_{-}|}{r+r_{-}}\right) }{3r_{-}^{2}+l^{2}-(l^{2}+r_{-}^{2})\mathcal{W}%
\left( \dfrac{l^{2}q^{2}}{r_{-}^{2}(l^{2}+r_{-}^{2})}\right) },
\label{eqtor1}
\end{equation}%
The point where the ingoing null rays from the two asymptotic
regions meet inside the black hole between the two horizons
($r_{-}<r_{meet}<r_{+}$) can be calculated numerically using Eqs.
(\ref{eqtor1})-(\ref{eqtor3}) which reads
\begin{equation}
r^{\ast }(r_{meet})=0.
\end{equation}%
\begin{center}
\begin{figure}[tbp]
\hspace{0.4cm} \centering
\subfigure{\includegraphics[width=0.4\columnwidth]{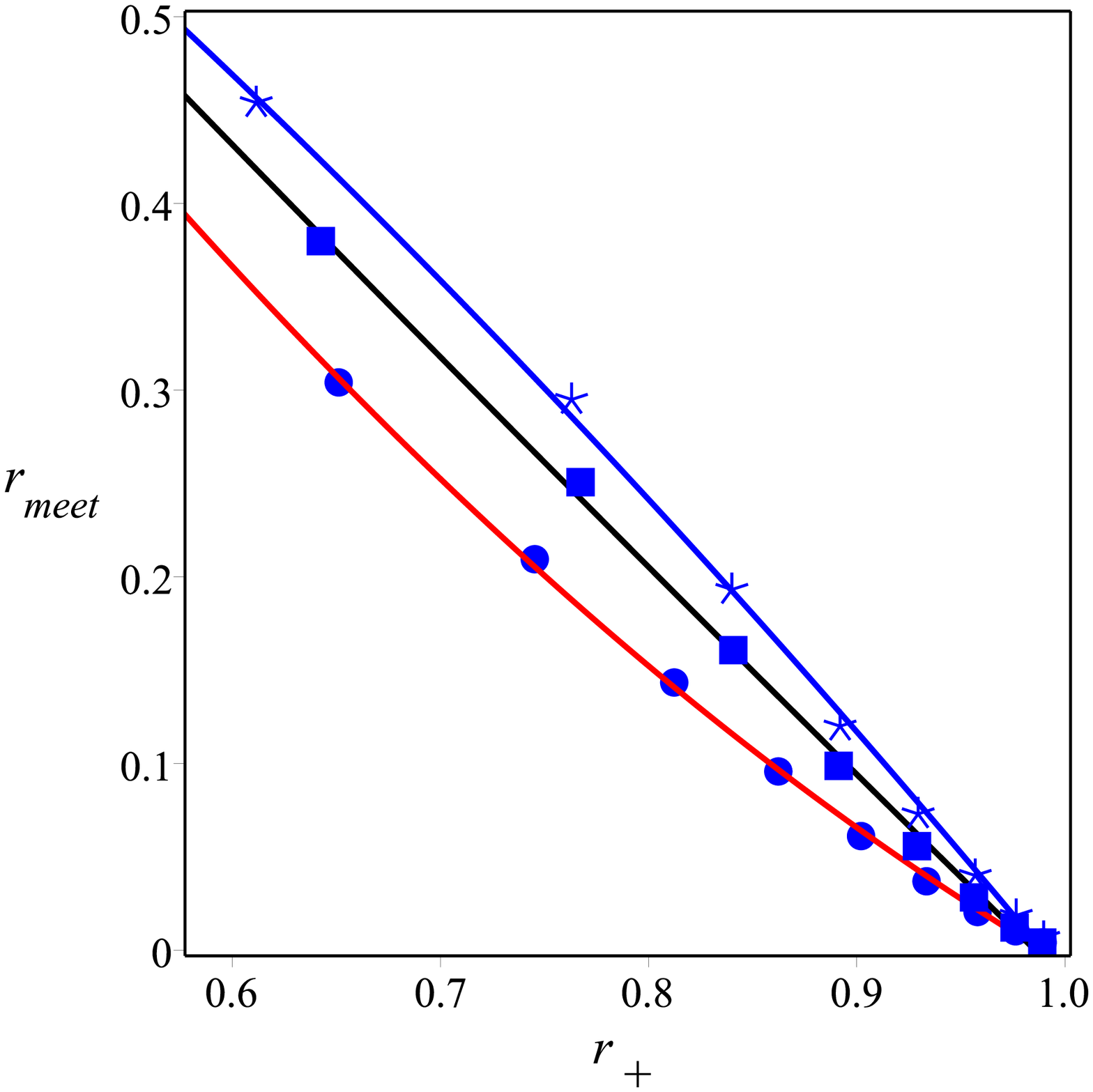}} %
\subfigure{\includegraphics[width=0.4\columnwidth]{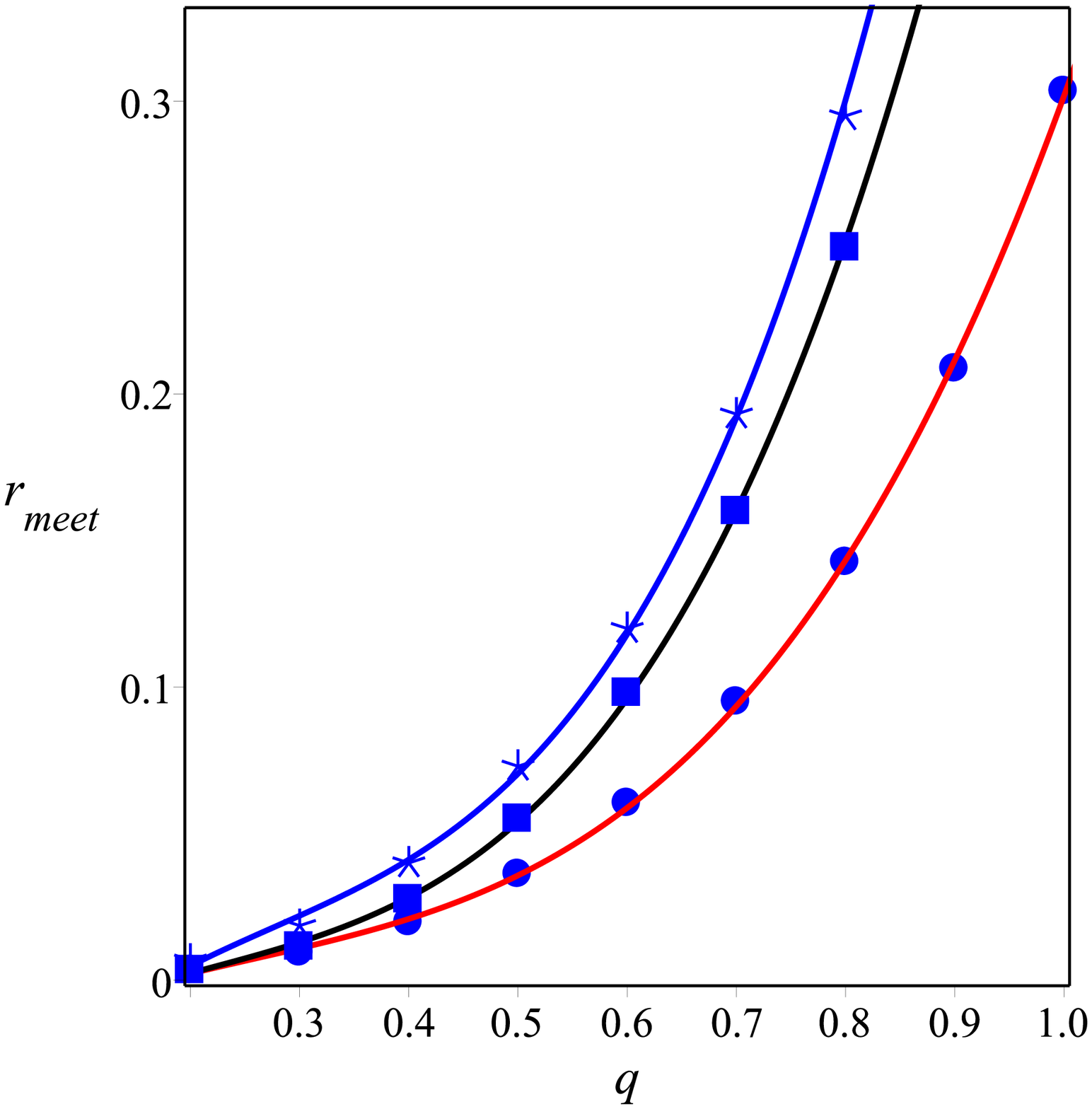}}
\caption{The behavior of $r_{meet}$ in terms of $r_{+}$ (left) and $q$
(right) for $M=l=1$. red line (first case), black line (second case) and
blue line (third case).}
\label{fig5}
\end{figure}
\end{center}
We show the results for $r_{meet}$ in Fig. \ref{fig5} for the considered
solution. According to the Fig. \ref{fig5}a, by increasing event horizon
radius, $r_{meet}$ decreases from extremal value to $r_{-}$. Then, by
increasing the electric charge, $r_{meet}$ increases (Fig. \ref{fig5}b). By
using of numerical curve fitting and choosing the appropriate branches of
the logarithms, one can obtain the following relation for $r_{meet}(r_{+})$
\begin{equation}
r_{meet}\approx 0.6825r_{+}^{2}-2.0259r_{+}+1.336,
\end{equation}
The above approximation functions are necessary to obtain the $A_{bulk}$ and
$A_{joint}$ in the following.\\
The bulk action of the black hole is
\begin{align}
A_{bulk,BH}=& 4\dfrac{4\pi }{16\pi G}\int_{r_{meet}}^{r_{max}}\left(
\mathcal{R}+\dfrac{6}{l^{2}}-\mathcal{L}\right) dr\int_{0}^{v_{\infty
}-r^{\ast }(r)}dt  \notag \\
=& \dfrac{2q^{2}}{G}\int_{r_{meet}}^{r_{max}}dr\left( \dfrac{\exp \left( -%
\dfrac{q^{2}}{2Mr}\right) }{r^{4}}\right) (v_{\infty }-r^{\ast }(r))\approx
\dfrac{8.27 r_{+}^2}{G l^4}-\dfrac{0.005 r_{+}}{G l^4}+\dfrac{0.00003 r_{+}}{%
G \delta l^2}+\mathcal{O}(1),
\end{align}%
where $v_{\infty }=\lim_{r\rightarrow \infty }r^{\ast }(r)$ is the
constant defining the future null boundary, $r_{max}=\frac{l^{2}}{\delta }-%
\frac{\delta }{4}+\frac{M\delta ^{2}}{3l^{2}}-\frac{q^{2}\delta ^{3}}{8l^{4}}
$ and $\delta $ is a some cutoff surface associated with a UV divergence. In
order to obtain the action of AdS spacetime, one uses%
\begin{equation}\label{eqads0}
f_{AdS}=1+\dfrac{r^{2}}{l^{2}},
\end{equation}%
in which the tortoise coordinate for AdS spacetime is%
\begin{equation}\label{eqads1}
r_{0}^{\ast }{}(r)=l\arctan \left( \dfrac{r}{l}\right) ,\hspace{0.5cm}%
v_{0\infty }=\dfrac{\pi l}{2}.
\end{equation}%
Thus the bulk action of AdS vacuum is%
\begin{align}\label{eqads2}
A_{AdS}& =4\dfrac{4\pi }{16\pi G}\int_{0}^{r_{0max}}dr\left( \mathcal{R}+%
\dfrac{6}{l^{2}}\right) \int_{0}^{v_{0\infty }-r_{0}^{\ast }{}(r)}dt=-\dfrac{%
6}{Gl^{2}}\int_{0}^{r_{0max}}dr(v_{0\infty }-r_{0}^{\ast }{}(r))=  \notag \\
& \dfrac{-1}{G}\left[ -l^{2}\ln \left( 1+\dfrac{r_{0max}^{2}}{l^{2}}\right)
+r_{0max}^{2}\left( 1+\dfrac{\pi r_{0max}}{l}-\dfrac{2r_{0max}}{l}\arctan (%
\dfrac{r_{0max}}{l})\right) \right] ,
\end{align}%
where $\mathcal{R}_{ads}=-12/l^{2}$ and $r_{0max}=\frac{l^{2}}{\delta }-%
\frac{\delta }{4}$. Now, we should calculate the following relation for the
solution
\begin{equation}
\Delta A_{bulk}=A_{bulk,BH}-A_{AdS}.  \label{eqact}
\end{equation}%
%
%
%
The action at the meet point $r_{meet}$ is%
\begin{equation}
\mathcal{A}_{joint}=\frac{2}{G}r_{meet}^{2}\ln [f(r_{meet})],  \label{eqjoin}
\end{equation}%
which in the case of large black hole becomes%
\begin{equation*}
\mathcal{A}_{joint}\approx 0.0186\ln \left( \dfrac{0.0682r_{+}^{2}}{l}%
\right) r_{+}^{4}+\mathcal{O}(r_{+}^{3}).
\end{equation*}%
The total action is the sum of the bulk (\ref{eqact}) and joint (\ref{eqjoin}%
) terms. Substituting the numerical solution for $r_{meet}$, we
obtain the interesting result plotted in Fig. \ref{CDV1}.
According to these figures, one finds the bulk action of black
hole and complexity of formation are linear functions of entropy.
For the large black hole, we have
\begin{equation}
\Delta \mathcal{C}_{\mathcal{A}}\approx \dfrac{8.27 C_{T}S}{l^{2}}+\mathcal{O%
}(S^{1/2}),
\end{equation}%
where $C_{T}=3l^{2}/\pi ^{3}G$ is the boundary central charge.

\begin{figure}[tbp]
\hspace{0.2cm} \subfigure{\includegraphics[width=0.7\columnwidth]{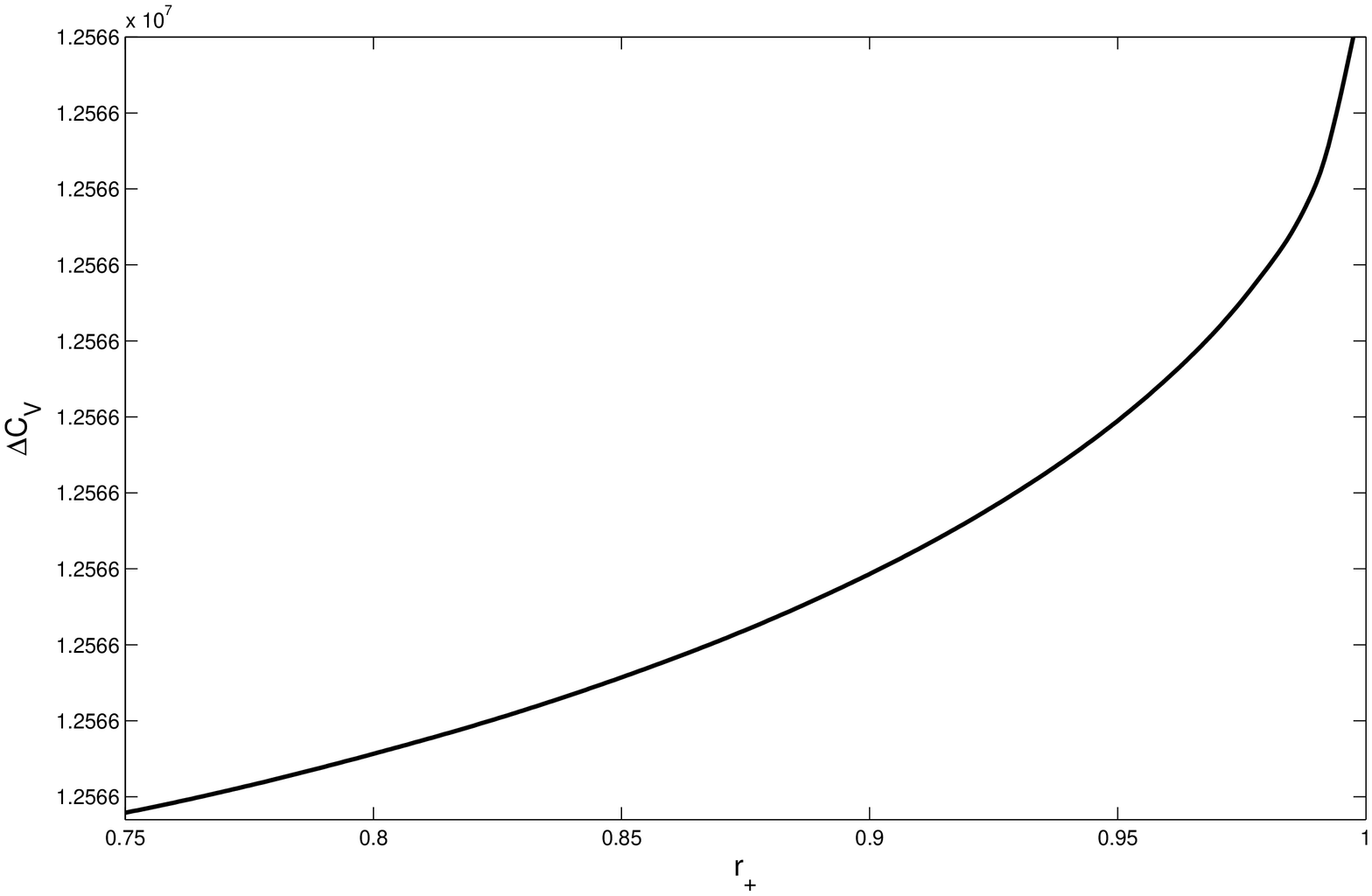}} %
\subfigure{\includegraphics[width=0.7\columnwidth]{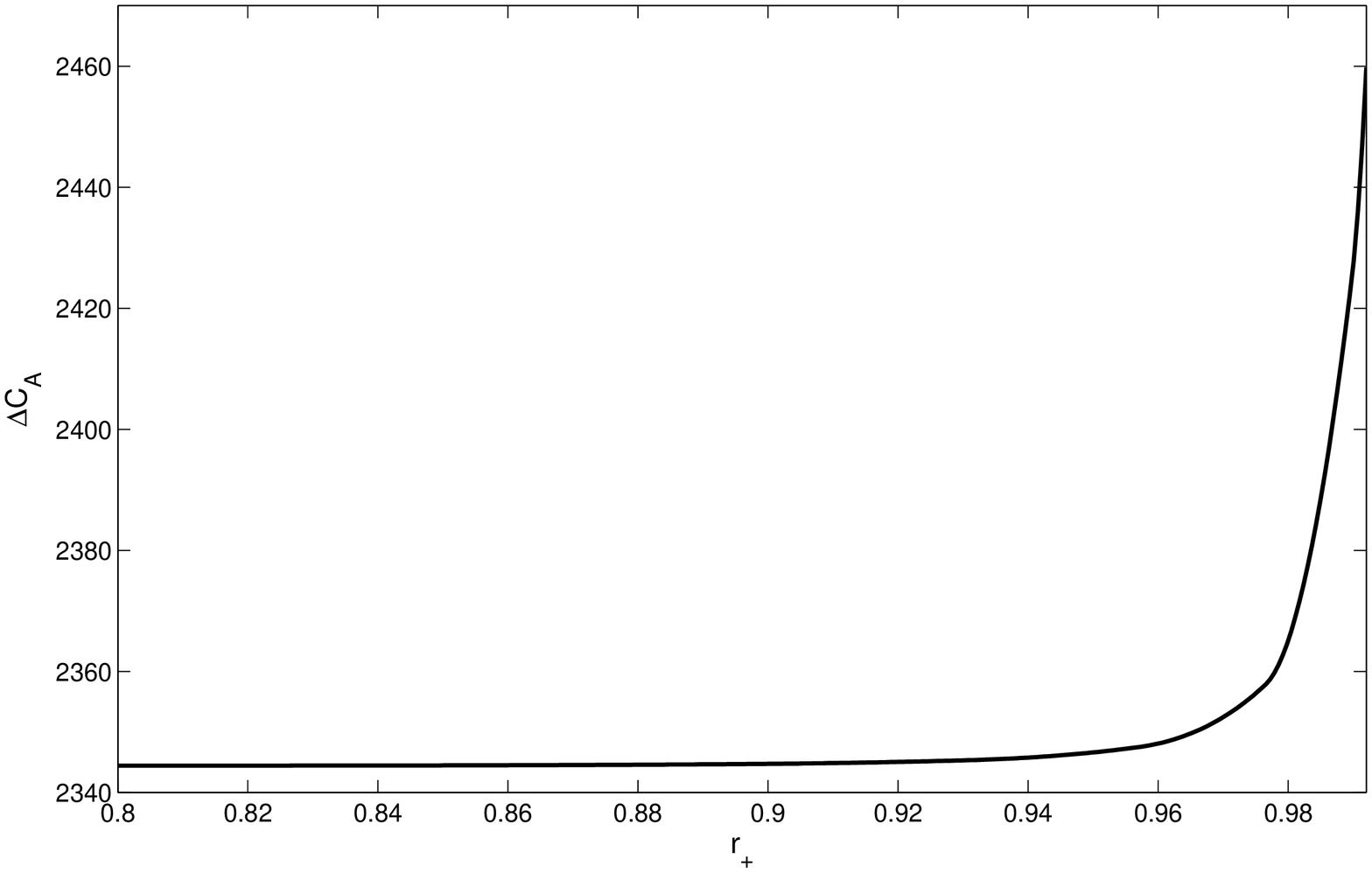}}
\caption{Plot of $\Delta \mathcal{C}_{\mathcal{V}} $ (up) and
$\Delta \mathcal{C}_{\mathcal{A}}$ (down) in terms of $r_{+}$ for
static exponential solution.. } \label{CDV1}
\end{figure}

Here, we want to obtain the complexity of formation by using of CV
conjecture. To do that, we consider the maximal volume functional for the $%
t=0$ timeslice (the straight line connecting the two boundaries through the
bifurcation surface in the Penrose diagram) as follows \cite%
{Chapman:2016hwi,Balushi:2020wkt}%
\begin{equation}
\mathcal{V}=8\pi \int_{r_{+}}^{r_{max}}dr\dfrac{r^{2}}{\sqrt{f}}=4\pi l
r_{+}^2+2\pi l^3+\dfrac{\pi l \delta^2}{4}+8\pi l^3 \ln\left(\dfrac{l}{\sqrt{%
\delta r_{+}}}\right)+\mathcal{O}\left(\dfrac{l^5}{r_{+}^5}\right),
\label{eqvolumfun}
\end{equation}%
where the corresponding contribution from two copies of the vacuum AdS
background is%
\begin{equation}
\mathcal{V}_{AdS}=8\pi \int_{0}^{r_{max}}dr\dfrac{r^{2}}{\sqrt{f_{AdS}}}%
=4\pi l\left[ r_{max}\sqrt{l^{2}+r_{max}^{2}}-l^{2}\ln \left( \dfrac{r_{max}+%
\sqrt{l^{2}+r_{max}^{2}}}{l}\right) \right] =\dfrac{4\pi l^{5}}{\delta ^{2}}%
-4\pi l^{3}\ln \left( \dfrac{2l}{\delta }\right) -\dfrac{\pi
l\delta ^{2}}{4}.
\end{equation}%
The complexity of formation is obtained by subtracting from (\ref{eqvolumfun}%
) the volume of the AdS vacuum
\begin{equation}\label{eqdeltacvform}
\Delta \mathcal{C}_{\mathcal{V}}=\dfrac{\mathcal{V}-2\mathcal{V}_{AdS}}{%
G_{N}R}=\dfrac{8\pi }{G_{N}R}\left[ \int_{r_{+}}^{r_{max}}\dfrac{r^{2}dr}{%
\sqrt{f(r)}}-\int_{0}^{r_{max}}\dfrac{r^{2}dr}{\sqrt{f_{AdS}}}\right] =%
\dfrac{41.3C_{T}S}{l^{3}}+\mathcal{O}(S^{1/2}).
\end{equation}%
To understand the behaviour of complexity of formation, we obtain
the plot of $\Delta \mathcal{C}_{\mathcal{V}}$ as a function of
$r_{+}$ in Fig. \ref{CDV1}.

It is interesting, of course, to compare the results of CA and CV
duality. The ratio of the complexity of formation for large black
holes $r_{+}\gg l$ is
\begin{equation}\label{Rform}
\mathcal{R}_{form}=\dfrac{\Delta \mathcal{C}_{\mathcal{A}}}{\Delta \mathcal{C%
}_{\mathcal{V}}}\approx 0.2,
\end{equation}%
and by using (\ref{Rrate}) and (\ref{Rform}) one gets
\begin{equation}
\mathcal{R}_{rate}-\mathcal{R}_{form}=0.02.
\end{equation}
As can be seen, the differences between two ratios is small. So, the two
ratios agree very well and this comparison suggests that the two holographic
approaches to complexity are consistent.

\section{Generalization to rotating solution:}\label{sec4}

The metric of rotating charged regular black hole in the Boyer-Lindquist
coordinates is obtained as
\begin{equation}
dS^{2}=-\dfrac{\Delta _{r}}{\Sigma }\left( dt-\dfrac{a\sin ^{2}(\theta )}{%
\Xi }d\phi \right) ^{2}+\dfrac{\Sigma }{\Delta _{r}}dr^{2}+\dfrac{\Sigma }{%
\Delta _{\theta }}d\theta ^{2}+\dfrac{\Delta _{\theta }\sin ^{2}(\theta )}{%
\Sigma }\left( adt-\dfrac{r^{2}+a^{2}}{\Xi }d\phi \right) ^{2},
\label{metric}
\end{equation}%
where
\begin{align}
\Delta _{r}& =(r^{2}+a^{2})(1+\dfrac{r^{2}}{l^{2}})-2f,\;\;\;\;\Sigma
=r^{2}+a^{2}\cos ^{2}(\theta ),\;\;\;\;\Xi =1-\dfrac{a^{2}}{l^{2}},  \notag
\\
\Delta _{\theta }& =1-\dfrac{a^{2}}{l^{2}}\cos ^{2}(\theta
),\;\;\;\;f(r)=Mr\exp \left( -\dfrac{q^{2}}{2Mr}\right) .
\end{align}%
The outer and inner horizons are determined by the equation $\Delta (r_{\pm
})=0$, respectively. The required thermodynamical quantities to check the
first law are \cite{Hendi:2020knv}%
\begin{align}
P=& \dfrac{3}{8\pi l^{2}},\hspace{0.5cm}V=\dfrac{r_{+}A}{3}+\dfrac{4\pi J^{2}%
}{3\mathcal{M}},\hspace{0.5cm}A=\dfrac{4\pi (r_{+}^{2}+a^{2})}{\Xi },\hspace{%
0.5cm}J=\dfrac{Ma}{\Xi ^{2}},\hspace{0.5cm}\mathcal{M}=\dfrac{M}{\Xi ^{2}},%
\hspace{0.5cm}Q=\dfrac{q}{\Xi },  \notag \\
T=& -\dfrac{(2Mr_{+}+q^{2})e^{-\left( \dfrac{q^{2}}{2Mr_{+}}\right) }}{4\pi
(r_{+}^{2}+a^{2})r_{+}}+\dfrac{r_{+}(r_{+}^{2}+2r_{+}^{2}+a^{2})}{2\pi
l^{2}(r_{+}^{2}+a^{2})},\hspace{0.5cm}\Phi =\dfrac{qr_{+}}{r_{+}^{2}+a^{2}},%
\hspace{0.5cm}\Omega =\dfrac{a}{r_{+}^{2}+a^{2}}\left( 1+\dfrac{r_{+}^{2}}{%
l^{2}}\right) .
\end{align}%
where
\begin{equation}
M=\dfrac{q^{2}}{2r_{+}W\left( \dfrac{l^{2}q^{2}}{%
(r_{+}^{2}+a^{2})(r_{+}^{2}+l^{2})}\right) }.
\end{equation}%
So, the first law of thermodynamic is satisfied as \cite{Hendi:2020knv}%
\begin{equation}
d\mathcal{M}=TdS+PdV+\Phi dQ+\Omega dJ,
\end{equation}%
Integrating the on-shell Einstein-Hilbert bulk action, directly,
in the case of $\frac{q}{M}\ll 1$ and $\frac{a}{l} \ll 1$ leads to
\begin{align}
\frac{d\mathcal{A}_{bulk}}{dt}& =\frac{1}{8}\int_{r_{-}}^{r_{+}}\int_{0}^{%
\pi }\sqrt{-g}\left( \mathcal{R}+\frac{6}{l^{2}}-\mathcal{L}\right)
drd\theta =  \notag \\
& -\left[ \dfrac{%
24Mr^{7}-24Ml^{2}q^{2}r^{3}+9l^{2}q^{4}r^{2}+24Ma^{2}r^{5}+24Ma^{2}q^{2}l^{2}r-11a^{2}l^{2}q^{4}%
}{48\Xi ml^{2}r^{4}}\right] _{r_{-}}^{r_{+}},
\end{align}%
since $\sqrt{-g}=\dfrac{\sin\theta }{\Xi}\Sigma$ and%
\begin{equation}
\mathcal{L}=\dfrac{2q^{2}e^{-\dfrac{q^{2}}{2Mr}}((Mr+q^{2})a^{4}\cos
^{4}\theta +(-6Mr+q^{2})a^{2}r^{2}\cos ^{2}\theta +Mr^{5})}{%
Mr(r^{2}+a^{2}\cos ^{2}\theta )^{4}}.
\end{equation}%
The growth rate of the surface term in the case of $\frac{q}{M}\ll
1$ and $\frac{a}{l}\ll 1$
(Eq. (\ref{eqI})) is%
\begin{equation}
\dfrac{d\mathcal{A}_{boundary}}{dt} =\frac{1}{8\pi }\int_{\partial \mathcal{%
M}}(\sqrt{-h}\mathcal{K})d\Omega _{2}=\left[ \dfrac{\Delta _{r}^{^{\prime }}%
}{4\Xi }\right] _{r_{-}}^{r_{+}}=
\left[ \dfrac{-8M^{2}r^{2}l^{2}+8Mr^{3}(l^{2}+a^{2}+2r^{2})+q^{4}l^{2}}{%
16Ml^{2}\Xi r^{2}}\right] _{r_{-}}^{r_{+}},
\end{equation}%
since the determinant of the induced metric on the null
hypersurface $r_{\pm }$ is $h=-\dfrac{\sin ^{2}\theta \Sigma
\Delta _{r}}{\Xi ^{2}}$ and the trace of extrinsic curvature can
be calculated from Eq. (\ref{extcur}) and the normal vector is
$n_{\mu }=(0,\sqrt{\frac{\Delta _{r}}{\Sigma }},0,0)$. Combining
the bulk action and boundary term, we can obtain the growth rate
of total action as follows
\begin{equation}
\dfrac{d\mathcal{A}}{dt}=\left[ \dfrac{%
24Mq^{2}l^{2}r^{3}-6q^{4}l^{2}r^{2}-24Ma^{2}q^{2}l^{2}r+11a^{2}q^{4}l^{2}-24M^{2}l^{2}r^{4}+24Ml^{2}r^{5}+24Mr^{7}%
}{48\Xi Ml^{2}r^{4}}\right] _{r_{-}}^{r_{+}},
\end{equation}%
where
\begin{equation*}
q^{2}=\dfrac{r_{+}(a^{2}l^{2}+a^{2}r_{-}^{2}+r_{-}^{2}l^{2}+r_{-}^{4})\ln
\left( \dfrac{r_{-}(l^{2}+r_{+}^{2})(a^{2}+r_{+}^{2})}{%
r_{+}(a^{2}+r_{-}^{2})(l^{2}+r_{-}^{2})}\right) }{(r_{+}-r_{-})l^{2}\left(
\dfrac{r_{-}(l^{2}+r_{+}^{2})(a^{2}+r_{+}^{2})}{%
r_{+}(a^{2}+r_{-}^{2})(l^{2}+r_{-}^{2})}\right) ^{\frac{r_{+}}{r_{-}-r_{+}}}}.
\end{equation*}
In the case of $\alpha \ll 1$, $\beta \ll 1$ and $\epsilon \ll 1$, one finds%
\begin{eqnarray}
q^{2}/r_{+}^{2} &=&\left( \dfrac{60.26}{\alpha ^{2}}-100.43+\left( -\dfrac{%
100.43}{\alpha ^{2}}+140.6\right) \beta ^{2}\right) +  \notag \\
&&\left( -\dfrac{120.51}{\alpha ^{2}}+100.43+\dfrac{100.43\beta ^{2}}{\alpha
^{2}}\right) \epsilon +\mathcal{O}(\epsilon ^{2})+\mathcal{O}(\alpha ^{2})+%
\mathcal{O}(\beta ^{4}),
\end{eqnarray}%
while for $\alpha \gg 1$, $\beta \ll 1$ and $\epsilon \rightarrow 0$, we have%
\begin{eqnarray}
q^{2}/r_{+}^{2} &=&2.72+\dfrac{13.59}{\alpha ^{2}}-\left( 8.15+\dfrac{29.9}{%
\alpha ^{2}}\right) \beta ^{2}-  \notag \\
&&\left( 2.72+\dfrac{27.2}{\alpha ^{2}}-\dfrac{29.9\beta ^{2}}{\alpha ^{2}}%
\right) \epsilon +\mathcal{O}(\epsilon )+\mathcal{O}(\beta ^{4})+\mathcal{O}%
(1/\alpha ^{4}).
\end{eqnarray}

In addition, in the case of $\alpha \ll 1$, $\beta \ll 1$ and $\epsilon
\rightarrow 0$, we obtain%
\begin{equation}
\dfrac{d\mathcal{A}}{dt}\approx \dfrac{22.3 r_{+}^3}{l^{2}}-\dfrac{9.32
r_{-}r_{+}^2}{l^2} +\mathcal{O}(r_{+}),
\end{equation}%
and for $\alpha \ll 1$, $\beta \ll 1$ and $\epsilon \rightarrow 1$, one
achieves%
\begin{equation}
\dfrac{d\mathcal{A}}{dt}\approx \dfrac{0.21r_{+}\alpha ^{4}\beta ^{6}}{%
\epsilon ^{12}}+\mathcal{O}(\epsilon ^{-11}),
\end{equation}%
since

\begin{align}
V/r_{+}^{3}\approx & \dfrac{0.8\times 10^{-9}\alpha ^{6}\beta ^{6}}{\epsilon
^{12}},\hspace{1cm}\epsilon \rightarrow 1,  \notag \\
V/r_{+}^{3}\approx & 84.13+560.9\alpha ^{6},\hspace{1cm}\epsilon \rightarrow
0.
\end{align}
As can be seen in both cases $\epsilon \rightarrow 1,0$, the rate of
complexity proportional to the volume of black holes instead of entropy of
black holes.

\subsection{Complexity of formation}

In this section complexity of formation for rotating black holes
in the limit $\frac{a}{l}\ll1$ and $\frac{q}{M}\ll1$ and in the
conjectures Complexity-Action and Complexity-Volume is calculated.

To do so, we calculated Lagrangian of the bulk and vacuum AdS in
the limit $ \frac{a}{l}\ll1$ and $\frac{q}{M}\ll1$, which in the
second-order approximation are written the following form
\begin{equation*}
L_{bulk}=\frac{-4r^{3}l^{2}-4r^{3}a^{2}-4a^{2}l^{2}r+\frac{%
4q^{2}l^{2}(l^{2}+a^{2})}{r}-\frac{4a^{2}q^{2}l^{4}}{r^{3}}}{8l^{4}},\;\;\;\;\;\;L_{AdS}=-\frac{2r^{3}}{l^{2}}.
\end{equation*}
For obtaining the complexity of formation we should compute the following relation:
\begin{equation*}
\pi \Delta C_{A}=A_{bulk}-A_{Ads}+A_{joints}=\int \left( \int \sqrt{-g}
\pounds _{bulk}(r_{\infty }^{\ast}-r^{\ast })dr-\int \sqrt{-g} \pounds %
_{AdS}(r_{\infty }^{\ast }-r^{\ast })dr\right) d\theta
+A_{joints},
\end{equation*}
where
\begin{equation*}
\pounds _{bulk}=R+\frac{6}{l^{2}}-L,\;\;\;\;\pounds
_{AdS}=R+\frac{6}{l^{2}},\;\;\;\;r^{\ast }=\int
\sqrt{\frac{g_{rr}}{g_{tt}}}dr.
\end{equation*}
Since $\lim_{r \rightarrow \infty }r^{\ast }=0\ $and$\ r^{\ast }$
in the limit $\frac{q}{M}\ll1$ and $\frac{a}{l}\ll 1$ can be
calculated analytically, but the integral of the first term with
respect to $r$ can not be calculated analytically thus it is
actually more convenient to use integration\ by parts to eliminate
the appearance of $r^{\ast }$ inside this expression, because
after that we must integrating with respect to $\theta $ therefore
the first integral must be calculated analytically,

\begin{equation*}
A_{bulk}=\int\int\sqrt{-g}\pounds _{bulk}(-r^{\ast})drd\theta =\int\left(
(-r^{\ast})\int\sqrt{-g}\pounds _{bulk} dr+\int\left(\sqrt{-\frac{g_{rr}}{%
g_{tt}}}\int \sqrt{-g}\pounds _{bulk}dr\right)dr \right) d\theta.
\end{equation*}

\begin{center}
\begin{figure}[tbp]
\centering
\includegraphics[width=14cm]{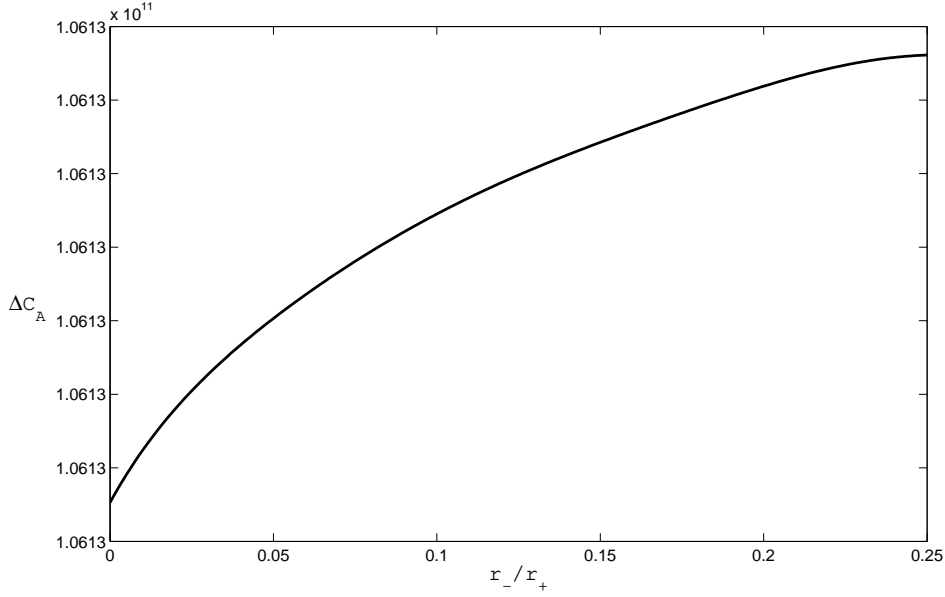}
\caption{Complexity of formation in the CA conjecture versus
$\frac{r_-}{r_+} $ for rotating black hole.} \label{dcasr}
\end{figure}
\end{center}

$A_{Ads}$ is obtained in the same way. After some calculations
complexity of formation for slowly rotating black holes in the
limit $\frac{q}{M}\ll 1$ within Complexity-Action conjecture is
obtained which is shown in Fig. \ref{dcasr}.

\begin{figure}[tbp]
\hspace{0.2cm}
\subfigure{\includegraphics[width=0.7\columnwidth]{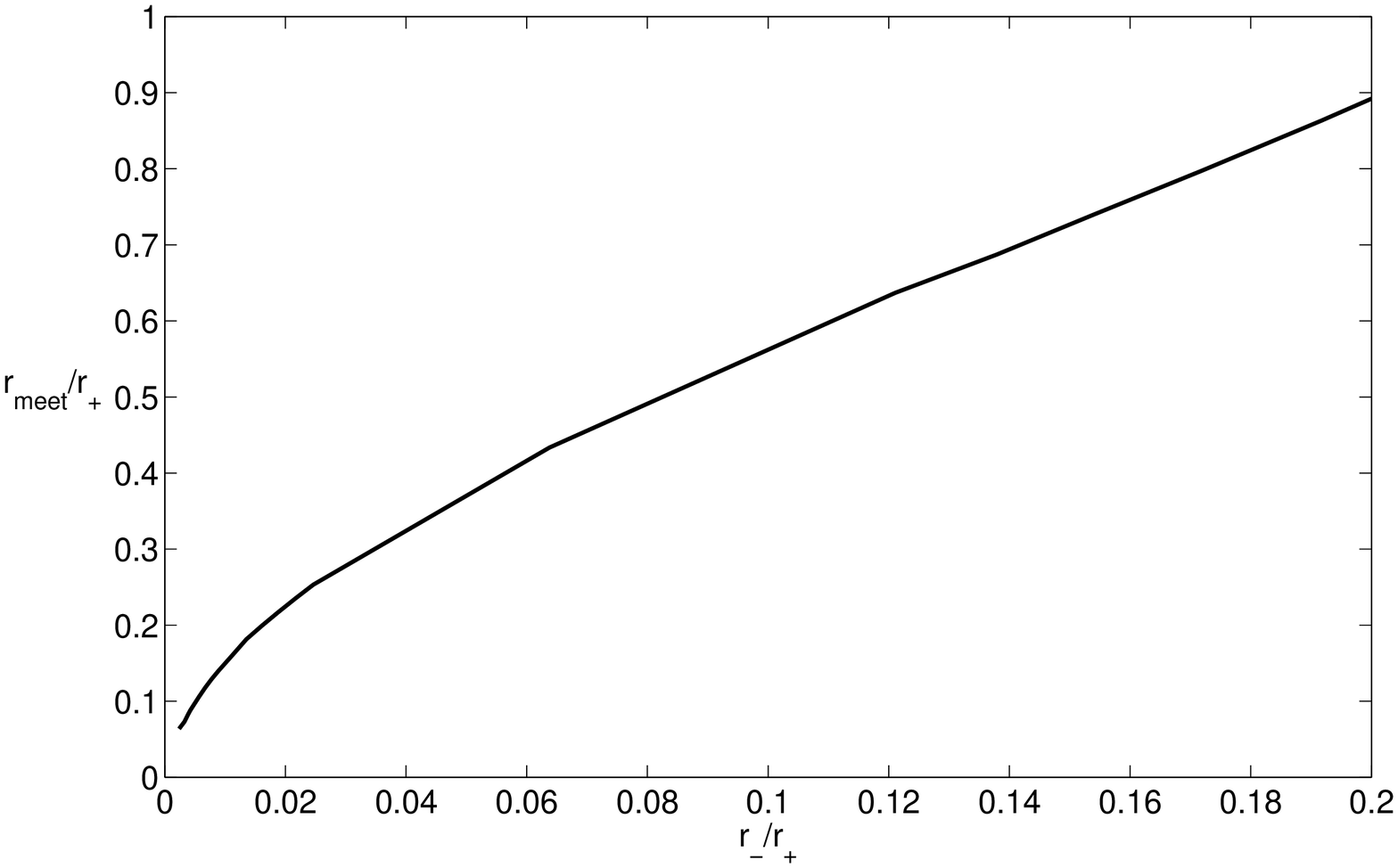}}
\subfigure{\includegraphics[width=0.7\columnwidth]{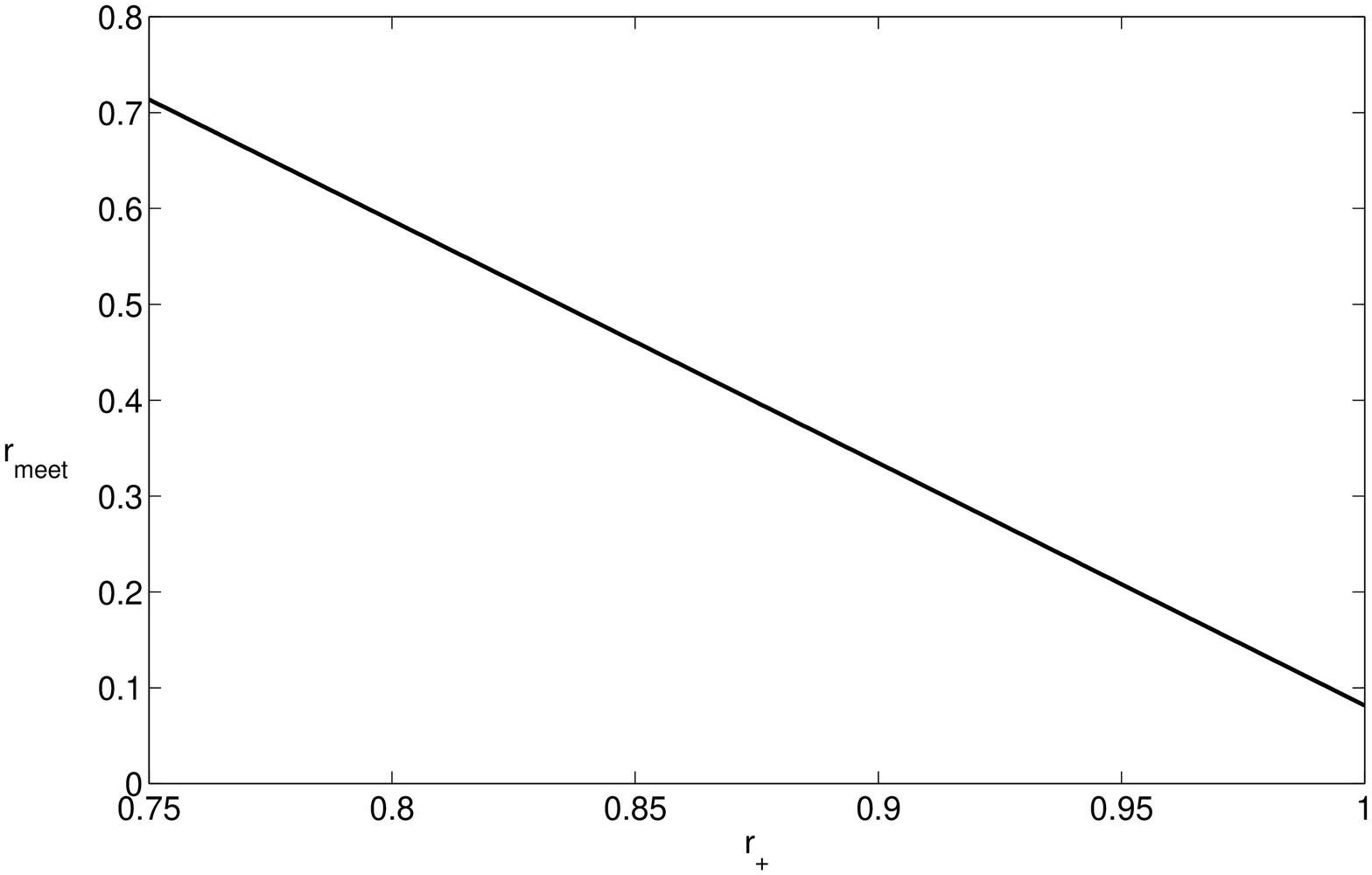}}
\caption{ The behavior of $\frac{r_{meet}}{r_{+}}$ versus
$\frac{r_{-}}{r_{+}}$ (up) and $r_{meet}$ versus $r_{+}$ (down).}
\label{meet12}
\end{figure}

The behavior of  $\frac{r_{meet}}{r_{+}}$ ($r_{meet}$) versus
$r_{-}/r_{+}$ ($r_{+}$) is plotted in Fig. \ref{meet12}. We find
that $r_{meet}$ is approximately a linear function of event
horizon and the best fitting curve relation is $r_{meet}=-2.5
r_{+} + 2.6$.


\textbf{Complexity of formation in CV conjecture}

For calculating the complexity in CV conjecture, we must find the
surface with maximal volume. For this purpose we should write the
line element in the tortoise coordinate
\begin{eqnarray*}
ds^{2} &=&d\upsilon ^{2}\left( -\frac{\Delta _{r}}{\Sigma }+\frac{%
a^{2}\Delta _{\theta }\sin ^{2}(\theta )}{\Sigma }\right) +dr^{2}\left( -%
\frac{\Delta _{r}}{\Sigma }\frac{g_{rr}}{g_{tt}}+\frac{a^{2}\Delta _{\theta
}\sin ^{2}(\theta )}{\Sigma }\frac{g_{rr}}{g_{tt}}+\frac{\Sigma }{\Delta _{r}%
}\right) \\
&&+2drd\upsilon \sqrt{-\frac{g_{rr}}{g_{tt}}}\left( \frac{\Delta _{r}}{%
\Sigma }-\frac{a^{2}\Delta _{\theta }\sin ^{2}(\theta )}{\Sigma }\right) +%
\frac{\Sigma }{\Delta _{\theta }}d\theta ^{2}+d\phi ^{2}\left( -\frac{\Delta
_{r}}{\Sigma }\frac{a^{2}\sin ^{4}(\theta )}{\Xi ^{2}}+\frac{%
(a^{2}+r^{2})^{2}\Delta _{\theta }\sin ^{2}(\theta )}{\Xi ^{2}\Sigma }\right)
\\
&&-d\phi d\upsilon \left( \frac{a\sin ^{2}(\theta )}{\Sigma \Xi }\right)
\left( 2\Delta _{r}-(a^{2}+r^{2})\Delta _{\theta }\right) +drd\phi \sqrt{-%
\frac{g_{rr}}{g_{tt}}}\left( -2\frac{\Delta _{r}}{\Sigma }+\frac{%
a(a^{2}+r^{2})\Delta _{\theta }\sin ^{2}(\theta )}{\Xi \Sigma
}\right),
\end{eqnarray*}
where
\begin{equation*}
\upsilon =t+r^{\ast },\;\;\;\;r^{\ast }=\int \sqrt{-\frac{g_{rr}}{g_{tt}}}dr=\int \sqrt{\frac{\Sigma ^{2}}{%
\Delta _{r}^{2}-a^{2}\Delta _{r}\Delta _{\theta }\sin ^{2}(\theta )}}dr.
\end{equation*}
To find the maximal surface we describe the surface with the
parametric relations $r=r(\lambda )$ and $\upsilon =\upsilon
(\lambda )$ with some parameter $\lambda$, then the volume of the
surface becomes in the following form
\begin{equation}\label{eqmaxvol}
V=\int \sqrt{\sigma }d^{2}x=4\pi \int \int dr d\theta r^{2}\left( \sqrt{\left[-%
\frac{\Delta _{r}}{\Sigma }+\frac{\Delta _{\theta }a^{2}\sin ^{2}(\theta )}{%
\Sigma }\right]\dot{\upsilon} ^{2}+2\dot{r}\dot{\upsilon} \sqrt{1-\frac{\Delta _{\theta }a^{2}\sin
^{2}(\theta )}{\Delta _{r}}}}\right),
\end{equation}
where dots indicate derivatives with respect to $\lambda $ and the
factor 2 is originated from this fact that the surface is composed
from two equivalent parts. We are free to choose $\lambda $ to
keep the radial volume element fixed as follow
\begin{equation}\label{eq415}
r^{2}\sqrt{\left[-\frac{\Delta _{r}}{\Sigma }+\frac{\Delta_{\theta }a^{2}\sin^{2}%
}{\Sigma}\right]\dot{\upsilon} ^{2}+2\dot{r}\dot{\upsilon}\sqrt{1-\frac{\Delta _{\theta }a^{2}\sin
^{2}(\theta) }{\Delta _{r}}}}=1,
\end{equation}
this Lagrangian is independent of $\upsilon $ and hence there is a
conserved quantity given by
\begin{equation}\label{eq416}
E=-\frac{\partial\pounds }{\partial \upsilon}=\left(\frac{\Delta_{r}}{ \Sigma}-%
\frac{\Delta_{\theta} a^{2}\sin (\theta)^{2}}{\Sigma}\right)\dot{\upsilon}-\sqrt{%
1-\frac{\Delta_{\theta} a^{2}\sin (\theta)^{2}}{\Delta_{r}}}\dot{r}.
\end{equation}

By substituting $\dot{\upsilon}$ from (\ref{eq415}) into the
(\ref{eq416}), one can obtain
\begin{equation*}
E^{2}=r^{4}\left(\frac{\Delta _{r}}{\Sigma }-\frac{\Delta _{\theta
}a^{2}\sin ^{2}(\theta )}{\Sigma
}\right)+r^{8}\dot{r}^{2}\sqrt{1-\frac{\Delta _{\theta }a^{2}\sin
^{2}(\theta )}{\Delta _{r}}},
\end{equation*}
with substituting above equations into the relation
(\ref{eqmaxvol}) for maximal surface volume one can find that
\begin{equation*}
V=4\pi \int_{r_{\min }}^{r_{\max}} \int_{0}^{\pi}\frac{dr}{\dot{r}}=2\Omega
\int_{r_{\min }}^{r_{\max }}\int_{0}^{\pi }\frac{r^{4}\sqrt[4]{1-\frac{%
\Delta _{\theta }a^{2}\sin ^{2}(\theta )}{\Delta _{r}}}d\theta dr}{\sqrt{%
r^{4}(\frac{\Delta _{r}}{\Sigma }-\frac{\Delta _{\theta }a^{2}\sin
^{2}(\theta )}{\Sigma })+E^{2}}}.
\end{equation*}

Here we wish to take $r_{\max }$ to be infinity, but this will
yield a divergent result in general. A finite result can be
obtained by performing a carefully matched subtraction of the AdS
vacuum. Here $r_{min}$\ is the turning point of the surface which
determined by the condition $\dot{r}=0$
\begin{equation*}
E^{2}+r_{\min }^{4}\left(-\frac{\Delta _{r}}{\Sigma }+\frac{\Delta
_{\theta }\sin ^{2}(\theta )}{\Sigma }\right)_{r_{\min }}=0.
\end{equation*}

A simple calculation shows that $r_{min}$ will be on or inside the (outer)
horizon, and so we have that, $g_{tt}(r_{\min })< 0$ , $\upsilon
(\lambda _{\min })> 0$ $\Longrightarrow $ E$< 0$ and we recall that $%
g_{tt}(r_{\min })< 0$ in the region between the inner and event
horizon. We are interested in maximal slice which $\tau =0$\ in
this case $r_{\min }=r_{+}$\ which gives $E=0$. Thus, the
complexity of formation becomes
\begin{eqnarray*}
\Delta C_{V} =\frac{V_{bulk}-2V_{ads}}{G_{N}L}=\frac{4\pi }{G_{N}L}\left(
\int_{r_{+}}^{r_{\max }}\int_{0}^{\pi }\frac{r^{2}\sqrt{\frac{\Sigma }{%
\Delta _{r}}}}{(1-\frac{\Delta _{\theta }a^{2}\sin ^{2}(\theta )}{\Delta _{r}%
})^{\frac{1}{4}}}d\theta dr-
 \int_{0}^{r_{0\max }}\int_{0}^{\pi }2r^{2}\sqrt{\frac{\Sigma }{%
r^{2}(1+\frac{r^{2}}{l^{2}})-2Mr}}d\theta dr\right),
\end{eqnarray*}
integrating with respect to $\theta$ gives us
\begin{eqnarray*}
\Delta C_{Vbulk} &=&\frac{8l\pi ^{2}}{G_{N}L}\int_{r_{+}}^{r_{\max }}\frac{dr}{%
\left[ r^{3}-(-r^{2}+2m)l^{2}\right] ^{\frac{5}{2}}}\left[ \left(\frac{r^{7}}{2}%
-r^{6}m+\left(-\frac{q^{2}}{8}-\frac{3}{32}a^{2}\right)r^{5}+\frac{5a^{2}q^{2}r^{3}}{32}%
\right)l^{2}+\frac{r^{9}}{4}-\frac{a^{2}r^{7}}{16}+\right.  \\
&&\left. \left(\frac{1}{4}r^{5}-mr^{4}+\left(-\frac{q^{2}}{8}+m^{2}-\frac{a^{2}}{32}%
\right)r^{3}+\left(\frac{q^{2}}{4}m-\frac{a^{2}m}{16}\right)r^{2}+\frac{a^{2}}{4}\left(m^{2}+\frac{%
7q^{2}}{16}\right)r+\frac{a^{2}q^{2}m}{16}\right)l^{4}\right]-\frac{8l\pi ^{2}}{G_{N}L}\times  \\
&& \int_{0}^{r_{0\max }}dr\frac{l\sqrt{r}}{%
\sqrt{(r-2m)l^{2}+r^{3}}}.
\end{eqnarray*}

\begin{figure}[tbp]
\hspace{0.4cm} \centering
\includegraphics[width=14cm]{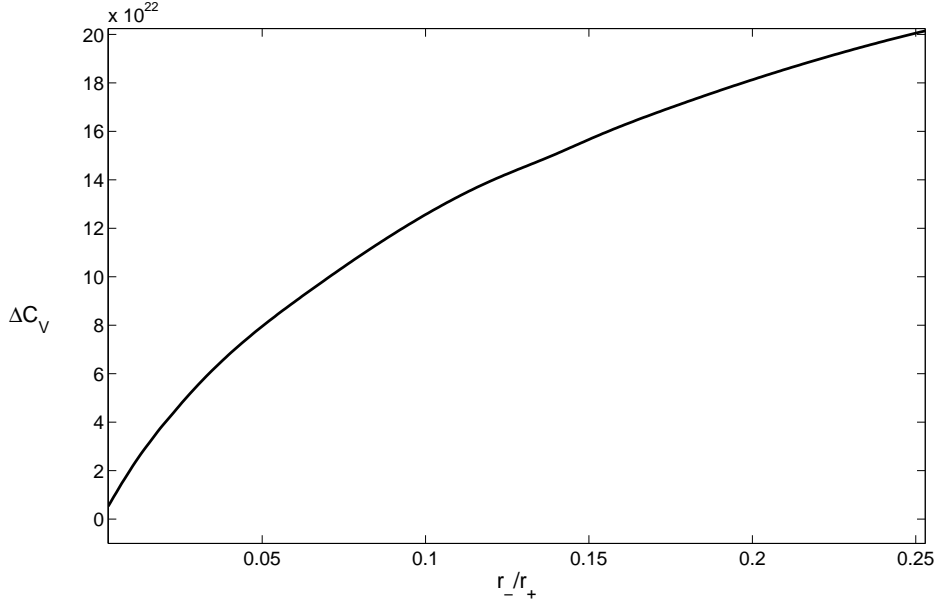}
\caption{Complexity of formation in the CV conjecture versus $\frac{r_-}{r_+}
$.}
\label{dcvsr}
\end{figure}

As the next step, we have to integrate the above relation with
respect to $r$. In this case the integral cannot be calculated
analytically, and therefore, we should calculate it numerically
which the results are shown in Fig. \ref{dcvsr}.

According to Figs. \ref{dcasr} and \ref{dcvsr}, it is obvious that
complexity of formation in two mentioned conjectures has same
behavior and it is bounded without divergency. In the limit
$\frac{r_-}{r_+}\rightarrow 0$ the value of the complexity of
formation (calculated via CV conjecture) is reduced and tends to
$4\times 10^{21}$ while at the large value of $\frac{r_-}{r_+}\sim
0.25$, it's value approximately tends to a constant. Since
increasing the electric charge leads decreasing (increasing) inner
horizon (event horizon), it is easy to find that $\frac{r_-}{r_+}$
is an increasing function of electric charge, $q$, and therefore,
the complexity of formation in both conjectures increases as one
increases the electric charge.

\section{Conclusion \label{sec5}}

In this paper, we studied the complexity of regular static and
rotating black holes in the presence of nonlinear electrodynamics
. For the case of the static black
hole, by looking at the action growth rates of the Wheeler-De Witt
patch, we showed the Lloyd bound is satisfied
and its form remains unaltered
. Then, we have obtained the rate of complexity and complexity of
formation by using both CA and CV conjectures. From the comparison
of the result, we concluded that the two approaches to the
complexity are consistent. In the case of slowly rotating one and
with assumption $\frac{q}{M}\ll 1$, we obtained the rate of
complexity proportional to the volume of the black hole (in the
limit $\alpha\ll 1$ and $\beta \ll1$). So, one can conclude that
the rate of complexity for both static and rotating black holes is
proportional to thermodynamical volume instead of entropy.

The complexity of formation is numerically obtained by using both
conjectures. Its behavior is reflected in Figs. \ref{dcasr} and
\ref{dcvsr} which shows that for both conjectures, the complexity
of formation has the same behavior and is bounded.

In the appendix, we investigated two other regular cases and
obtained the rate of complexity and complexity of formation by
using of both conjectures for them. According to Figs. \ref{dcast}
and \ref{dcvst}, it is obvious that for each case, the complexity
of formation obtained from two approaches are relatively
consistent and on the other hand behavior of complexity of
formation for these two regular cases have interesting
similarities with each other.


\textbf{Acknowledgements}\\
The authors thank Shiraz University Research Council.

\section{Appendix}
For the sake of completeness, here, we investigate complexity
conjecture of two more examples of static regular black holes in
$\mathcal{P}$.

\subsection{Second Case:}

Here, we consider another structure function inspired by the
log-logistic distribution as the second class of solution which is
given by \cite{Balart:2014cga,Dymnikova:2004zc}
\begin{equation}
\mathcal{H}=\frac{\mathcal{P}}{(1+\gamma_{0} \sqrt{-\mathcal{P}})^{2}}+\frac{6}{%
l^{2}},  \label{H2}
\end{equation}%
where $\gamma_{0} =\frac{r_{0}^{2}}{\sqrt{2\gamma ^{2}}}$, and
$r_{0}$ and $\gamma
$ are two parameters. In the weak field limit ($\mathcal{P}\ll 1$), Eq. (\ref%
{H2}) reduces to
\begin{equation}
\mathcal{H}\approx \mathcal{P}+\frac{6}{l^{2}}+\mathcal{O}(\mathcal{P}^{%
\frac{3}{2}}).
\end{equation}

Taking such a structure function into account with
$\mathcal{H}=2\rho (r)$ and Eqs. (\ref{eqn:p}) and
(\ref{eqmetric1}), we can, directly, achieve the following
$arctan$ form of the solution
\cite{Dymnikova:2004zc,Sajadi:2017glu}
\begin{equation}
f(r)=1-\frac{4\chi }{\pi r}\left[ arctan(x)-\frac{x}{1+x^{2}}\right] +%
\frac{r^{2}}{l^{2}},\;\;\;r_{0}=\frac{\pi \gamma ^{2}}{8\chi },\;\;\;x=%
\frac{r}{r_{0}},
\end{equation}
with the following asymptotical behavior for large $r$
\begin{equation}
-g_{tt}\approx 1+\frac{r^{2}}{l^{2}}-\frac{2\chi }{r}+\frac{\gamma ^{2}}{%
r^{2}}+\mathcal{O}\left( \frac{1}{r^{4}}\right).
\end{equation}%
As a result, the obtained solution behaves like
Reissner-Nordstr\"{o}m-AdS, asymptotically, when we adjust $\chi $
and $\gamma $ as the mass and electric charge of the system,
respectively.

Such as before, we can obtain the following thermodynamic
quantities for the recent $arctan$ solution
\begin{eqnarray}
P &=&\frac{3}{8\pi l^{2}}, \\
V &=&\frac{4\pi r_{+}^{3}}{3}, \\
S &=&\pi r_{+}^{2}, \\
T &=&\frac{m\left( (1+x_{+}^{2})^{2}\arctan
(x_{+})-x_{+}-3x_{+}^{3}\right)
}{\pi ^{2}r_{+}^{2}(1+x_{+}^{2})^{2}}+\frac{r_{+}}{2\pi l^{2}}, \\
\Phi &=&\int_{r_{+}}^{\infty }Edr=\frac{3m}{\pi q}\left( \frac{\pi }{2}%
-\arctan (x_{+})+\frac{(1+\frac{5}{3}x_{+}^{2})x_{+}}{(1+x_{+}^{2})^{2}}%
\right) ,
\end{eqnarray}%
where $x_{+}=x|_{r=r_{+}}$. Having the mentioned thermodynamic
quantities, one can directly examine the first law of black hole
thermodynamics in the enthalpy representation
\begin{equation}
dH=TdS+VdP+\Phi dq,
\end{equation}
where
\begin{equation}
H=m=\frac{\pi r_{+}(1+\frac{r_{+}^{2}}{l^{2}})}{4\left( \arctan (x_{+})-%
\frac{x_{+}}{1+x_{+}^{2}}\right) }.
\end{equation}

Moreover, the Smarr relation is given by
\begin{equation}
\frac{H}{2}+PV-TS-\frac{q\Phi }{2}+\frac{1}{4}\int wdv=0,
\end{equation}%
where
\begin{equation}
\int wdv=\frac{1}{2}\int_{r_{+}}^{\infty }T_{\mu }^{\mu }\ r^{2}dr=m-\frac{%
2mx_{+}(x_{+}^{2}-1)}{\pi (1+x_{+}^{2})^{2}}-\frac{2m}{\pi
}\arctan (x_{+}).
\end{equation}

Now, we calculate joint terms according to joint contribution
which is described in the first class. After simplification, we
can write
\begin{eqnarray}
{S_{B^{\prime }B}} &=&\frac{1}{4}\delta t\ \left[ {\frac{{4m}}{{\pi {r^{2}}}}%
\left( {\arctan (x)-\frac{{x}}{1+x^{2}}}\right) -\frac{{4m}}{{\pi r}}\left( {%
\frac{{2x{^{2}}}}{{{{{r_{0}}(1+x^{2})}^{2}}}}}\right) +\frac{{2r}}{{{l^{2}}}}%
}\right] _{r_B},  \notag \\
{S_{C^{\prime }C}} &=&\frac{1}{4}\delta t\left[ {\frac{{4m}}{{\pi {r^{2}}}}%
\left( {\arctan (x)-\frac{{x}}{1+x^{2}}}\right) -\frac{{4m}}{{\pi r}}\left( {%
\frac{{2x{^{2}}}}{{{{{r_{0}}(1+x^{2})}^{2}}}}}\right) +\frac{{2r}}{{{l^{2}}}}%
}\right] _{r_C}.
\end{eqnarray}

Considering the metric function obtained here, it is easy to show
that the Ricci scalar is calculated as
\begin{equation}
\mathcal{R}=\frac{4q^{2}}{r_{0}^{4}(1+x^{2})^{3}}-\frac{12}{l^{2}},
\end{equation}%
and also, the Lagrangian of nonlinear electrodynamics can be
calculated at the event horizon with the following explicit form
\begin{equation}
\mathcal{L}=\frac{16m(1-x_{+}^{2})}{\pi r_{0}^{3}(1+x_{+}^{2})^{3}}-\frac{6}{%
l^{2}}.
\end{equation}

Consequently, the growth rate of bulk action and surface term are,
respectively,
\begin{equation}
\frac{d\mathcal{A}_{bulk}}{dt}=\left[ \frac{2m}{\pi }\left( \arctan (x)-{%
\frac{{x}}{1+x^{2}}}\right) \right] _{r_{-}}^{r_{+}},
\end{equation}
and
\begin{equation}
\frac{d\mathcal{A}_{boundray}}{dt}=\left[ r-\frac{3m}{\pi }\arctan (x)+\frac{%
mx(3+x^{2})}{\pi (1+x^{2})^{2}}+\frac{3r^{3}}{2l^{2}}\right]
_{r_{-}}^{r_{+}}.
\end{equation}

Finally, we find that the total growth rate of the action for such
black hole configuration within WDW patch at late time
approximation can be
collected as%
\begin{equation}
\frac{d\mathcal{A}}{dt}=\left[ r-\frac{m}{\pi }\arctan (x)-\frac{mx(1-x^{2})%
}{\pi (1+x^{2})^{2}}+\frac{3r^{3}}{2l^{2}}\right]
_{r_{-}}^{r_{+}}.
\end{equation}%
Thus, by using the above calculated thermodynamic quantities, it
is easy to
show that the Lloyd bound is satisfied. As an example, for the values $%
m=0.8,q=0.1,l=1$ this bound is about $1.6872$, and the differences
between left and right (right mines left of Eq. (\ref{rate})) is
about $0.0666$.

\subsection{Third Case:}

As the third case, we consider the following structure function
which is inspired by the Fermi-Dirac distribution
function\cite{AyonBeato:1999rg,Balart:2014cga}
\begin{equation}
\mathcal{H}=\frac{\mathcal{P}}{\cosh ^{2}\left( \frac{\gamma ^{\frac{3}{2}}(-%
\mathcal{P})^{\frac{1}{4}}}{2^{\frac{5}{4}}\chi }\right)
}+\frac{6}{l^{2}}, \label{H3}
\end{equation}%
with the following weak field limit ($\mathcal{P}\ll 1$)%
\begin{equation}
\mathcal{H}\approx \mathcal{P}+\frac{6}{l^{2}}+\mathcal{O}(\mathcal{P})^{%
\frac{3}{2}}.
\end{equation}


Considering the spherically symmetric spacetime with previous relations for $%
\mathcal{H}$, the third class of solution is%
\begin{equation}
f(r)=1-\frac{2\chi }{r}\left[ 1-\tanh \left( \frac{\gamma ^{2}}{2\chi r}%
\right) \right] +\frac{r^{2}}{l^{2}}. \label{f3}
\end{equation}%
According to the series expansion of the metric function for large
$r$, one can find this solution behaves like the
Reissner-Nordstr\"{o}m-AdS, asymptotically, as
\begin{equation}
-g_{tt}\approx 1+\frac{r^{2}}{l^{2}}-\frac{2\chi }{r}+\frac{\gamma ^{2}}{%
r^{2}}+\mathcal{O}\left( \frac{1}{r^{4}}\right),
\end{equation}%
whereas $\chi $ and $\gamma $ will be associated with the mass and
electric charge of the system, respectively.

Using the previous approach in the extended phase space, the
following thermodynamical quantities can be obtained, directly,
\begin{eqnarray}
P &=&\frac{3}{8\pi l^{2}}, \\
V &=&\frac{4\pi r_{+}^{3}}{3}, \\
S &=&\pi r_{+}^{2}, \\
T &=&\frac{m}{2\pi r_{+}^{2}}\left[ 1-\tanh \left( \frac{q^{2}}{2mr_{+}}%
\right) \right] -\frac{q^{2}}{4\pi r_{+}^{3}}\left[ 1-\tanh ^{2}\left( \frac{%
q^{2}}{2mr_{+}}\right) \right] +\frac{r_{+}}{2\pi l^{2}}, \\
\Phi &=&\int_{r_{+}}^{\infty }Edr=\frac{3m}{2q}-\frac{3mr_{+}\left( 1+e^{%
\frac{q^{2}}{mr_{+}}}\right)
-q^{2}e^{\frac{q^{2}}{mr_{+}}}}{qr_{+}\left(
1+e^{\frac{q^{2}}{mr_{+}}}\right) ^{2}}.
\end{eqnarray}%
It is evident that these thermodynamic quantities satisfy the
first law of black hole thermodynamics as
\begin{equation}
dH=TdS+VdP+\Phi dq,
\end{equation}%
where in the enthalpy representation, we have%
\begin{equation}
H=m.
\end{equation}%
In addition, the related Smarr formula is given by%
\begin{equation}
\frac{H}{2}+PV-TS-\frac{q\Phi }{2}+\frac{1}{4}\int wdv=0,
\end{equation}%
where
\begin{equation}
\int wdv=\frac{1}{2}\int_{r_{+}}^{\infty }T_{\mu }^{\mu }r^{2}dr=m-\frac{%
2mr_{+}\left( 1+e^{\frac{q^{2}}{mr_{+}}}\right) +2q^{2}e^{\frac{q^{2}}{mr_{+}%
}}}{r_{+}\left( 1+e^{\frac{q^{2}}{mr_{+}}}\right) ^{2}}.
\end{equation}%
According to the previous section, joint contribution can be
calculated as
below%
\begin{equation}
S_{B^{\prime }B}=\frac{1}{4}\delta t\left( {\frac{{2m}}{{r^{2}}}}\left[ {{%
1-\tanh }\left( {\frac{{q^{2}}}{{2mr}}}\right) }\right] {-\frac{{q^{2}}}{{%
r^{3}}}}\left[ {{1-{{{\tanh }^{2}\left( {\frac{{q^{2}}}{{2mr}}}\right) }}}}%
\right] {+\frac{{2r}}{{l^{2}}}}\right) _{r_B},
\end{equation}%
\begin{equation}
{S_{C^{\prime }C}}=\frac{1}{4}\delta t\left( {\frac{{2m}}{{r^{2}}}}\left[ {{%
1-\tanh }\left( {\frac{{q^{2}}}{{2mr}}}\right) }\right] {-\frac{{q^{2}}}{{%
r^{3}}}}\left[ {{1-{{{\tanh }^{2}\left( {\frac{{q^{2}}}{{2mr}}}\right) }}}}%
\right] {+\frac{{2r}}{{l^{2}}}}\right) _{r_C}.
\end{equation}%
Moreover, the Ricci scalar for this solution is simplified as%
\begin{equation}
\mathcal{R}=\frac{q^{4}\sinh \left( \frac{q^{2}}{2mr}\right)
}{mr^{5}\cosh ^{3}\left( \frac{q^{2}}{2mr}\right)
}-\frac{12}{l^{2}},
\end{equation}%
and the nonlinear Lagrangian is as follows
\begin{equation}
\mathcal{L}=\frac{q^{2}\left( q^{2}\tanh \left(
\frac{q^{2}}{2mr}\right)
-2mr\right) }{mr^{5}\cosh ^{2}\left( \frac{q^{2}}{2mr}\right) }-\frac{6}{%
l^{2}}.
\end{equation}%
It is notable that the growth rate of the bulk action and its
related
surface term can be written as%
\begin{eqnarray}
\frac{d\mathcal{A}_{bulk}}{dt} &=&\left[ \frac{2m}{1+e^{\left( \frac{q^{2}}{%
mr}\right) }}\right] _{r_{-}}^{r_{+}}, \\
\frac{d\mathcal{A}_{boundray}}{dt} &=&\left[ r-(\frac{3m}{2}+\frac{q^{2}}{4r}%
)+\frac{3m}{2}\tanh \left( \frac{q^{2}}{2mr}\right)
+\frac{q^{2}}{4r}\tanh ^{2}\left( \frac{q^{2}}{2mr}\right)
+\frac{3r^{3}}{2l^{2}}\right] _{r_{-}}^{r_{+}}.
\end{eqnarray}%
As a result, the total growth rate of action for such a black hole
configuration within WDW patch at late time approximation is%
\begin{equation}
\frac{d\mathcal{A}}{dt}=\left(
r+\frac{3r^{3}}{2l^{2}}-\frac{m}{2}\left[ 1-\tanh \left(
\frac{q^{2}}{2mr}\right) \right] -\frac{q^{2}}{4r}\left[ 1-\tanh
^{2}\left( \frac{q^{2}}{2mr}\right) \right] \right)
_{r_{-}}^{r_{+}}.
\end{equation}

For instance, for the values $m=0.8,q=0.1,l=1$ this bound is about
$8.5783$,
and the differences between left and right (right mines left of Eq. (\ref%
{rate})) is $7.0399$.

\subsection{Complexity of formation}

Similar to the complexity of formation of the first case we would
like to study the complexity of formation of second and third
cases. In this way, $r^{\ast }(r)$ and $r_{meet}$ of the second
and the third solutions can be obtained as follows:
\begin{align}
r^{\ast }(r)=& \dfrac{\pi l^{2}r_{+}^{2}\ln \left( \dfrac{|r-r_{+}|}{r+r_{+}}%
\right) }{4ml^{2}\arctan (\dfrac{r_{+}}{r_{0}})+\dfrac{2\pi r_{+}^{5}}{%
r_{+}^{2}+r_{0}^{2}}+\dfrac{2r_{0}r_{+}(\pi
r_{0}^{3}r_{+}^{2}-6ml^{2}r_{+}^{2}-4ml^{2}r_{0}^{2})}{%
(r_{+}^{2}+r_{0}^{2})^{2}}}+  \notag \\
& \dfrac{\pi l^{2}r_{-}^{2}\ln \left( \dfrac{|r-r_{-}|}{r+r_{-}}\right) }{%
4ml^{2}\arctan (\dfrac{r_{-}}{r_{0}})+\dfrac{2\pi r_{-}^{5}}{%
r_{-}^{2}+r_{0}^{2}}+\dfrac{2r_{0}r_{+}(\pi
r_{0}^{3}r_{-}^{2}-6ml^{2}r_{-}^{2}-4ml^{2}r_{0}^{2})}{%
(r_{-}^{2}+r_{0}^{2})^{2}}}\hspace{2.7cm}second\;case  \label{eqtor2} \\
r^{\ast }(r)=& \dfrac{l^{2}r_{+}^{3}\ln \left( \dfrac{|r-r_{+}|}{r+r_{+}}%
\right) }{2mr_{+}l^{2}\left( 1-\tanh \left(
\dfrac{q^{2}}{2mr_{+}}\right) \right) -q^{2}l^{2}\left( 1-\tanh
^{2}\left( \dfrac{q^{2}}{2mr_{+}}\right)
\right) +2r_{+}^{4}}+  \notag \\
& \dfrac{l^{2}r_{-}^{3}\ln \left( \dfrac{|r-r_{-}|}{r+r_{-}}\right) }{%
2mr_{-}l^{2}\left( 1-\tanh \left( \dfrac{q^{2}}{2mr_{-}}\right)
\right) -q^{2}l^{2}\left( 1-\tanh ^{2}\left(
\dfrac{q^{2}}{2mr_{-}}\right) \right)
+2r_{-}^{4}}\hspace{2.5cm}third\;case  \label{eqtor3}
\end{align}
and
\begin{align}
r_{meet}& \approx 0.068r_{+}^{2}-1.23r_{+}+1.1433,\hspace{1.5cm}%
second\;\;case \\
r_{meet}& \approx -0.35387r_{+}^{2}-0.64296r_{+}+0.9821.\hspace{0.5cm}%
third\;\;case
\end{align}
Accordingly, the bulk action of the black holes are
\begin{align}
A_{bulk,BH}=& \dfrac{16Mr_{0}}{\pi
G}\int_{r_{meet}}^{r_{max}}dr\left(
\dfrac{1}{(r^{2}+r_{0}^{2})^{2}}\right) (v_{\infty }-r^{\ast }(r))\hspace{1.5cm}second\;case \\
A_{bulk,BH}=& \dfrac{2q^{2}}{G}\int_{r_{meet}}^{r_{max}}dr\left( \dfrac{%
1-\tanh ^{2}\left( \dfrac{q^{2}}{2Mr}\right) }{r^{4}}\right)
(v_{\infty 5}-r^{\ast }(r))\hspace{0.5cm}third\;case
\end{align}
here $r_{max}$ is the same as the first case. The relation for
$A_{joint}$ for large black holes are written in the following
form:
\begin{align}
\mathcal{A}_{joint}\approx & 0.0185\ln \left(
\dfrac{0.068r_{+}^{2}}{l}\right)
r_{+}^{4}+\mathcal{O}(r_{+}^{3}).\;\;\;second\;case \\
\mathcal{A}_{joint}\approx & 0.1252\ln \left( \dfrac{-0.35387r_{+}^{2}}{l}%
\right) r_{+}^{4}+\mathcal{O}(r_{+}^{3}).\;\;\;third\;case
\end{align}

\begin{center}
\begin{figure}[tbp]
\centering
\includegraphics[width=12cm] {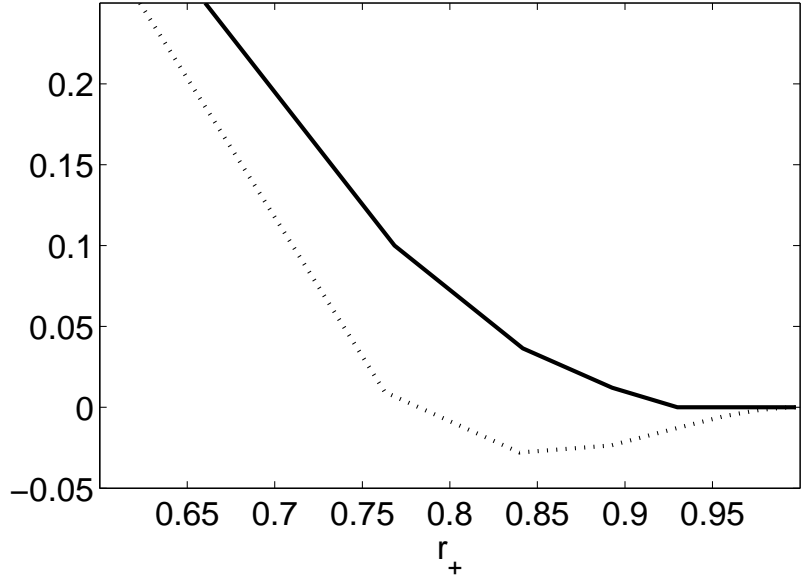}
\caption{Plot of $A_{joint}$ for second and third cases which are
shown in black and dotted line respectively.} \label{jointt}
\end{figure}
\end{center}

The figure of $A_{joint}$ for second and third cases are plotted
in Fig. \ref{jointt}. According to Fig. \ref{jointt}, it is
obvious that for small black hole the contribution of the joints
to the action is large for large black holes, this contribution
becomes smaller and goes to zero.

\begin{center}
\begin{figure}[tbp]
\centering
\includegraphics[width=12cm] {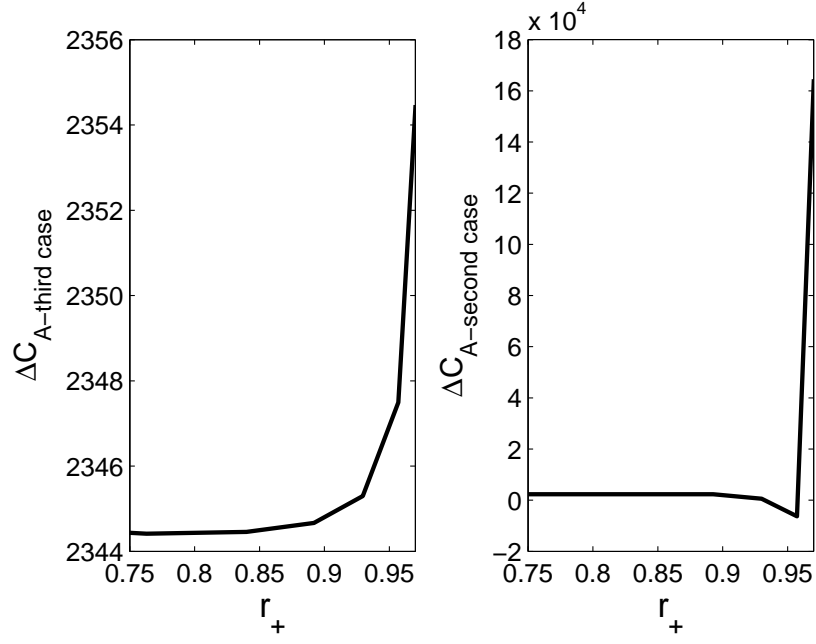}
\caption{Plots of $\Delta C_{A}$ for second and third cases are
shown in the right and left panels respectively.} \label{dcast}
\end{figure}
\end{center}

Finally according to above relations and the results related to
the $AdS$ spacetime (\ref{eqads0})-(\ref{eqads2}), $\Delta C_{A} $
from (\ref{eqdeltaca}) can be obtained. The plot of $\Delta C_{A}$
for second and third cases are shown in Fig. \ref{dcast}.

Figure \ref{dcast} describes that complexity of formation for
second and third cases increases for large black holes. Finally
$\Delta C_{V}$ according to Eq. (\ref{eqdeltacvform}) for these
two cases can be obtained.

\begin{center}
\begin{figure}[tbp]
\centering
\includegraphics[width=12cm] {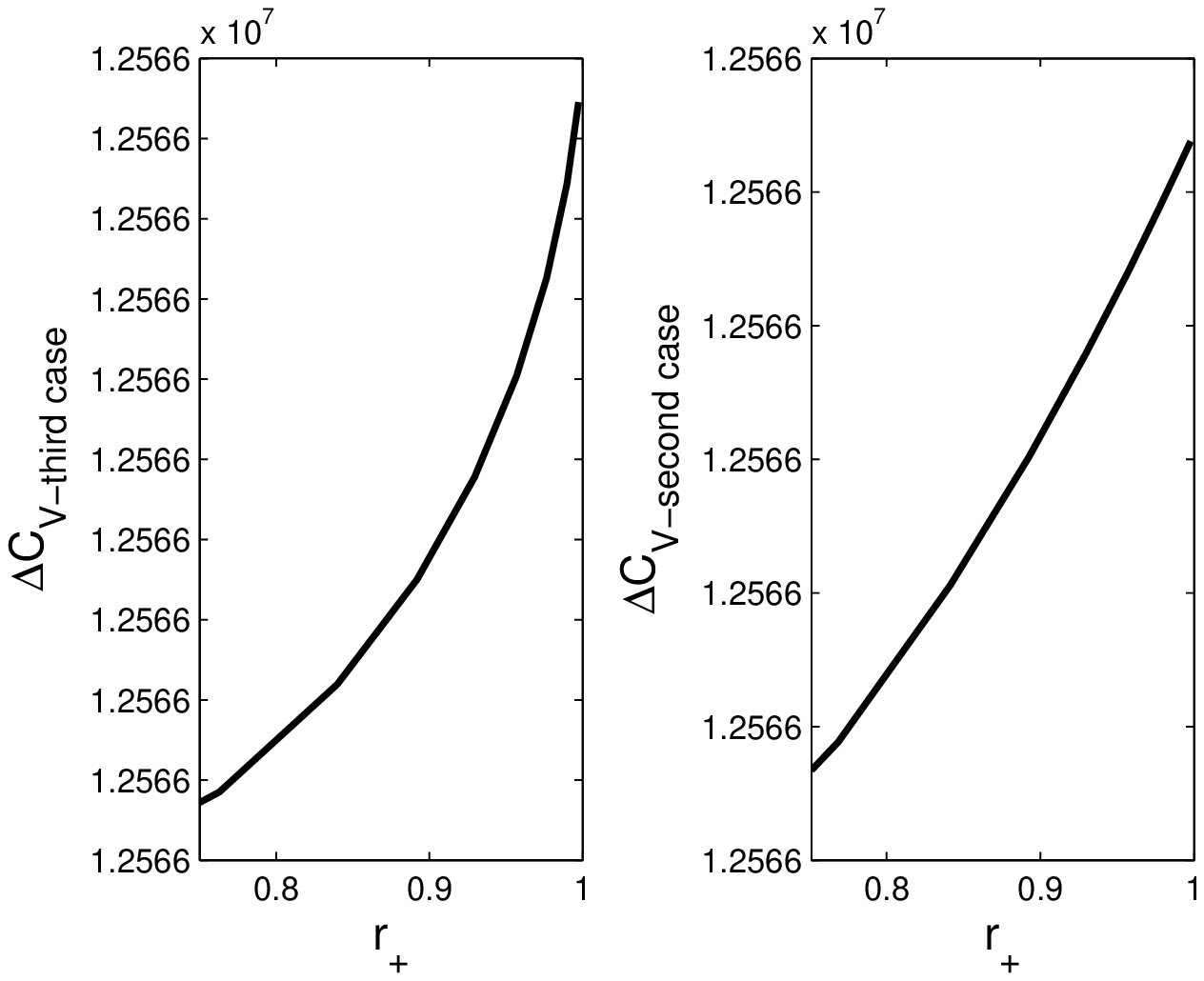}
\caption{Plots of $\Delta C_{V}$ for second and third cases are
shown in the right and left panels respectively.} \label{dcvst}
\end{figure}
\end{center}

\begin{center}
\begin{figure}[tbp]
\centering
\includegraphics[width=16cm] {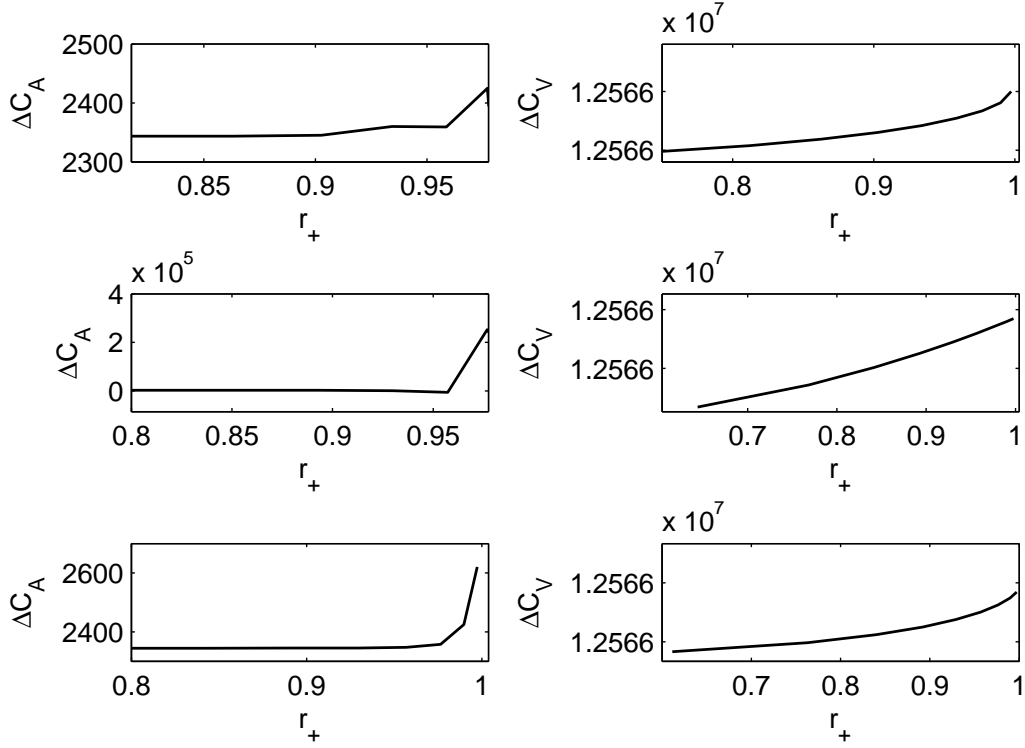}
\caption{Plot of Complexity of formation for the different
geometries. For first, second and third cases respectively are
shown from up to down, and left panels describe $\Delta C_{A}$ and
right ones indicate $\Delta C_{V} $.} \label{three}
\end{figure}
\end{center}

Figure \ref{dcvst} shows that $\Delta C_{V}$ for both cases has
the same behavior as with $\Delta C_{A}$, increasing with radius
of the black hole. To summarize the behavior of $\Delta C_{A}$ and
$\Delta C_{V}$ for three cases versus $r_{+}$ are plotted in Fig.
\ref{three}.



\end{document}